\documentclass[Afour,sageh,times]{sagej}

\usepackage{moreverb,url}
\usepackage{graphicx}
\UseRawInputEncoding
\usepackage{fvextra}
\usepackage{amsmath}
\usepackage{fancyvrb}
\usepackage{xcolor}
\usepackage{longtable}
\usepackage{comment}
\usepackage{alltt}  % 

\usepackage{booktabs,tabularx,threeparttable,graphicx}
\usepackage{siunitx}
\usepackage{tcolorbox}
\usepackage{tabularx}
\tcbuselibrary{listings,breakable,skins}

\newtcblisting{llmprompt}{
  breakable,
  enhanced,
  colback=white,
  colframe=black!20,
  boxrule=0.5pt,
  arc=2pt,
  left=6pt,right=6pt,top=6pt,bottom=6pt,
  listing only,
  listing options={
    basicstyle=\ttfamily\footnotesize,
    breaklines=true,
    breakatwhitespace=false,
    columns=fullflexible
  }
}

\usepackage[colorlinks,bookmarksopen,bookmarksnumbered,citecolor=red,urlcolor=red]{hyperref}

\newcommand\BibTeX{{\rmfamily B\kern-.05em \textsc{i\kern-.025em b}\kern-.08em
T\kern-.1667em\lower.7ex\hbox{E}\kern-.125emX}}

\def\volumeyear{2026}

\begin{document}

\runninghead{Liu et al.}

\title{Teacher-Authored Prompts for Configuring Student–AI Dialogue: K–12 Classroom Implementation}

\author{Alex Liu\affilnum{1}, Min Sun\affilnum{1}, Lief Esbenshade\affilnum{1}, Victor Tian\affilnum{1}, Zachary Zhang\affilnum{2}, and Kevin He\affilnum{2}}

\affiliation{\affilnum{1}Univeristy of Washington\\
\affilnum{2}Colleague AI, INC.}

\corrauth{Alex Liu, Univeristy of Washington
College of Education,
AmplifyLearn AI Center,
Seattle, WA,
98195, US.}

\email{alexliux@uw.edu}

\begin{abstract} GenAI has rapidly entered instructional and learning settings as a teaching assistant or AI tutor. However, less is known about how pedagogical intent connects to the learning generated within these systems, especially when student-facing AI dialogues are fine-tuned through teacher orchestration in live classrooms. This study examines a classroom deployment of a ``Classroom Teaching Aide'' (TASD) system, which enables teachers to author both a teacher-to-AI setup prompt (instructional scaffold) and a student-facing conversation starter to launch AI-mediated classroom discussions. We analyze a multi-subject pilot conducted in Spring 2025, involving 20 participating teachers (16 of whom implemented the system), across 39 classrooms and 77 TASD settings, yielding 1,479 student-AI conversations with 878 unique students. Using platform logs, LLM coding with human validation, and post-study teacher interviews (N=10), we characterize teacher authoring choices and link them to enacted student-AI interaction outcomes. Teachers predominantly authored highly specific tasks targeting higher-order thinking, with 92\% at Depth of Knowledge (DOK) Levels 2--3 \cite{webb2002}. In deployment, student-AI conversations were largely aligned with instructional intent: 71\% were fully on-track, and fewer than 1\% were substantially off-track. However, a persistent design-enactment gap emerged for cognitive demand: 38\% of conversations under-reached the teacher-targeted DOK level, approaching 50\% when targeting DOK 3. The study also shows that explicit finish lines in the prompt reduced the DOK gap by 0.22 levels (p $<$ .001), and ``no direct answers'' guardrails reduced AI final-answer rates by 8.5 percentage points. These findings position teacher-authored prompt layers as critical orchestration levers that translate pedagogical intent into structured student-AI dialogue, underscoring both their promise for scalable classroom integration and the need for additional supports to reliably sustain higher-order reasoning during enactment.
\end{abstract}

%%
%% Keywords. The author(s) should pick words that accurately describe
%% the work being presented. Separate the keywords with commas.
\keywords{Generative AI, Classroom Orchestration, Teacher-Authored AI Chat, K-12, Multilingual Learners, Personalized Learning at Scale, Human-AI Complementarity}
%% A "teaser" image appears between the author and affiliation
%% information and the body of the document, and typically spans the
%% page.
%%
%% This command processes the author and affiliation and title
%% information and builds the first part of the formatted document.
\maketitle

\section{Introduction}

Large language models (LLMs) have made conversational support broadly accessible, accelerating interest in AI tools that assist educators and students at scale \cite{kasneci2023}. In current educational use, AI is commonly configured as (a) a backstage teacher assistant for planning and materials generation \cite{moundridou2024, dennison2025}, (b) a teaching assistant surfaced through teacher-facing dashboards that suggest instructional moves based on classroom activity or tutoring-system data \cite{aleven2016a, holstein2019}, or (c) a direct-to-student learning platform in which the primary interaction occurs between learner and system with comparatively limited teacher control \cite{letourneau2025, vanlehn2011}. Each configuration reassigns roles between teachers, students, and AI systems, but ultimately it should be teachers who possess the professional expertise and contextual information to determine whether and how AI contributes to instructionally meaningful learning \cite{Lawrence2024}.

Classroom orchestration refers to teachers' real-time coordination of students, activities, tools, and instructional decisions to sustain coherent progress toward learning goals under practical constraints \cite{doyle1986, prieto2019}. In live classrooms, teachers must manage time, norms, safety, and alignment to curricular goals under real-time conditions \cite{leek2024}. When an additional AI agent is introduced into this workflow, teachers need to coordinate across students, tools, and teaching moments while sustaining pedagogical intent, potentially increasing managerial and cognitive load as they synthesize and translate output from multiple planes into actionable next steps \cite{jin2025}. Prior orchestration research emphasizes that classroom technologies are most viable when they incorporate direct support for coordination work across simultaneous learners and activities \cite[for eg.]{dillenbourg2013design, tissenbaum2019}. From this perspective, AI systems designed for synchronous classroom use must be legible, bounded, and supervisable in ways that foreground teachers' ability to configure, interpret, and oversee learning occurring within the technology, rather than emphasizing solely on underlying model capabilities \cite{roschelle2013, littleton2012}.

Prior research has demonstrated how AI-powered tools can support scalable teaching and learning, such as orchestration layers in MOOCs \cite{haaklev2017} or real-time awareness features in intelligent tutoring systems \cite{brunskill2024}. At the same time, studies of teacher-AI argumentation highlight that instructional value arises not from model fluency alone, but from how educators and AI systems coordinate roles, information, and decisions in the flow of classroom activity \cite{holstein2019, holstein2022}. Still, much of this work focuses either on enabling teaching or on shaping learning, with fewer investigations into how teacher can configure AI systems to mediate the connection between pedagogical intent and student learning within those systems \cite{unesco2023, jin2025, warschauer2023}.

This study examines teacher-student-AI facilitation through a teacher-authored student-AI dialogue tool that treats the teacher-authored prompt as an instructional design lever for shaping student discourse at classroom scale. The implemented system, Colleague AI Classroom Teaching Aide (TASD), enables a teacher-configured, student-facing AI conversation that preserves teacher intent while supporting individualized dialogue across students. A teacher authors (i) an AI-facing shared prompting layer that conditions the AI's role, discussion topics, scaffolding moves, and interaction constraints for the whole class or specific groups (e.g., multilingual learners, advanced students) and (ii) a common student-facing opening instruction that launches the discussion activity. Rather than treating AI chat as an autonomous tutor, TASD frames student-AI interaction as a configurable classroom routine that scales teacher-designed dialogue structures that preserve teachers' instructional intent and oversight. This approach aligns with AI system designs that rely on system prompts to guide and stabilize human-LLM interactions \cite{han2023, wang2024mathchat}, examining how these systems are enacted, configured, and orchestrated within real K-12 classroom conditions.

Guided by two research questions:
\begin{itemize}
    \item \textbf{RQ1 (Teacher authoring):} What do teachers build when given control to author and configure student-AI dialogues at classroom scale?
    \item \textbf{RQ2 (Classroom orchestration outcomes):} How do teacher-authored designs shape what actually happens in student-AI conversations, specifically task alignment, cognitive demand realization, student interaction moves, and AI scaffold adherence?
\end{itemize}
We conducted a multi-site pilot study in Spring 2025 to examine classroom deployment and interaction outcomes. Data sources included teacher-authored prompts, full student-AI conversation traces (including instances of non-use), and semi-structured teacher interviews. Together, these support an analysis of the full instructional chain: from teacher-authored instructional frames and entry scaffolds $\rightarrow$ to student participation and student-AI interactions $\rightarrow$ to outcomes such as task alignment, realization of cognitive demand, and adherence to AI use safeguards, interpreted through teachers' accounts of orchestration strategies and constraints. 

To address these questions, we conducted a multi-site pilot study during Spring 2025 across 16 teachers and 878 students. The study examines the full instructional chain: from teacher-authored instructional frames and entry scaffolds $\rightarrow$ to student participation and student-AI interactive outcomes. Drawing on descriptive and qualitative analysis of platform logs, student-AI conversations, and teacher interviews, we analyze how instructional intent is realized in practice. Findings highlight high task alignment, but a consistent gap between intended and actual cognitive demand, alongside evidence that prompt features such as finish lines and safety guardrails influence both student engagement and AI behavior. This study contributes to learning-at-scale and AI-in-education research by offering an empirical account of student-AI dialogue at classroom scale that foregrounds the connection between teachers' instructional intentions and student learning within AI systems, reinforcing transparency and teacher agency by making AI behavior controllable and interpretable to teachers.

The remainder of this paper proceeds as follows. We situate the study in prior work on classroom orchestration, human-AI complementarity, prompt-based interaction design, and student-facing conversational systems; describe TASD and the teacher authoring workflow; detail the pilot context and analytic approach; report results organized by research question; and conclude with implications, limitations, and directions for future research.

\section{Related Work}

This study is built up three interrelated strands of research: (1) how classroom technologies support real-time orchestration and learning at scale, (2) how task framing and discourse norms shape student participation and cognitive demand, and (3) how prompting functions as an interaction design mechanism in educational applications of GenAI. We bring these threads together by treating teacher-authored prompts as orchestration tools, designed not only to guide student-AI interactions but to preserve pedagogical intent, structure participation, and support cognitive rigor under authentic classroom conditions.

\subsection{Classroom Orchestration and Teacher-Facing Support at Scale}

Classroom orchestration describes how teachers coordinate multi-layered activity in real time, including pacing, participation, discipline, and alignment of tools and social arrangements to instructional goals \cite{dillenbourg2013design, dillenbourg2018}. Classroom technologies sometimes may even intensify orchestration load by introducing new decision points and monitoring demands, especially when many learners are active simultaneously \cite{ortega-arranz2024}. Learning-at-scale research has therefore emphasized abstractions and teacher-facing representations that make complex activity tractable with tools like orchestration graphs and real-time dashboards \cite{haaklev2017, tissenbaum2019}.

In AI-supported classrooms, teacher-facing awareness and intervention supports are repeatedly positioned as prerequisites for instructional integrity. Co-designed orchestration tools suggest that teacher-AI complementarity depends on actionable visibility and teacher-in-the-loop control, rather than autonomous system behavior alone \cite{holstein2019, holstein2022}. This human-AI co-constructive perspective emphasizes that AI should augment teacher capabilities rather than replace teacher judgment, providing recommendations, surfacing patterns, or handling routine interactions while teachers retain authority over instructional decisions \cite{baker2021, alam2023}.

\subsection{Conversational Agents for Learning}

Conversational agents for learning have a long history, from ELIZA-style interactions through dialogue-based intelligent tutoring systems demonstrating that structured dialogue can support explanation, reasoning, and learning \cite{wang2024eliza}. Meta-analyses and systematic reviews report that intelligent tutoring systems produce meaningful learning gains, particularly when dialogue is structured around productive pedagogical moves such as prompting for explanation, providing feedback on reasoning, and adapting to student knowledge states \cite{dong2025}.

Reviews of educational chatbots report growing adoption but uneven grounding in learning theory and frequent reliance on self-reports rather than rigorous outcome measures, highlighting that instructional value depends on task design and implementation conditions \cite{winkler2018, wollny2021, kuhail2023}. In language learning, systematic reviews find increased practice opportunities and engagement, with substantial variability by scaffolding, learner characteristics, and task structure \cite{huang2022, liu2025chatbotefl}. The consistent message is that conversational technology is not inherently effective; outcomes depend on how interaction is designed and implemented.

GenAI-based learning systems amplify these interaction design stakes because unconstrained dialogue can drift, overload learners with verbosity, or collapse into answer-seeking \cite{baidoo2023, yan2024practical}. Work such as RECIPE illustrates how structured guidance and role-setting can stabilize learner-LLM interaction toward pedagogically meaningful moves \cite{han2023}. Studies of students using ChatGPT for learning reveal both productive uses (explanation seeking, feedback on drafts) and problematic patterns (copying, over-reliance on AI responses) that depend on how interactions are framed and constrained \cite{denny2024, famaye2023ban}.

\subsection{Task Design, Discourse Quality, and Enacted Cognitive Demand}

Teachers' framing of discussion tasks has long been shown to shape how students participate, whether dialogue becomes brief answer exchanges or collaborative reasoning \cite{cazden1988discourse, michaels2008talk}. Norms like ``justify your answer'' or ``build on a peer's idea'' serve as participation contracts that elevate the quality of talk \cite{resnick2015talk, mercer2004talk}. Moment-to-moment prompts, such as revoicing, pressing for reasoning, or inviting elaboration, help sustain rigor and coherence during execution \cite{gillies2019, alexander2020}.

The link between design and thinking is well documented in the Depth of Knowledge (DOK) framework \cite{webb2002}: tasks designed for higher cognitive demand are more likely to elicit strategic or extended thinking, though implementation mediates this relationship \cite{stein1996, henningsen1997}. Previous studies consistently highlight the ``decline'' problem, occurring when rigor weakens during enactment despite high initial task design \cite{boston2012}. Structured participation routines can mitigate this gap and support equitable engagement \cite{sedova2019, asterhan2025}, especially when discussion norms are made explicit in the instructional framing \cite{xu2020wechat}.

\subsection{Prompting as Interaction Design for GenAI}

A growing body of work treats prompting not only as a technical mechanism for steering LLM outputs, but as an \emph{interaction design} practice that shapes what end users do, how assistance is sequenced, and whether conversational activity remains aligned to instructional intent \cite{zamfirescu2023johnny, white2024chatgpt}. In this framing, prompt choices redistribute responsibility among teacher, student, and AI system: they specify the AI's stance (e.g., coach versus answer engine), the expected epistemic work (e.g., justify, cite evidence, revise), and the completion criteria that turn open-ended chat into a bounded instructional routine \cite{holstein2019, kasneci2023, unesco2023}. In education-oriented dialogue settings, Socratic prompting has been explored as a mechanism for shifting the AI from answer-giving to question-asking, reinforcing the claim that prompt-level decisions encode an epistemic stance that can either support or undermine productive student participation \cite{liu2024socraticlm, orynbassarova2024}. Emerging scholarship further frames prompt engineering as a practice competency for educators, a skill to effectively align prompts to learning goals, constraints, and classroom norms \cite{walter2024}.

Recent GenAI learning systems operationalize this idea through \emph{prompt layers} and structured guidance intended to increase reliability and pedagogical alignment. For example, RECIPE demonstrates how a hidden instruction layer can set an instructional role and couple it with structured guidance aligned to writing goals, shaping learner-LLM interaction toward instructionally meaningful behaviors \cite{han2023}. MathChat and similar systems show how carefully designed prompts can scaffold mathematical reasoning through structured multi-turn dialogue \cite{wang2024mathchat}. Complementary research in prompt programming proposes reusable strategies: explicit role specification, decomposition into steps, requirements for justification, and use of exemplars, which increase controllability and reduce drift across turns \cite{reynolds2021prompt, white2024chatgpt, liu2023prompting}. 

Prompting also functions as a mechanism for responsible-use design. Guidance for GenAI in education consistently emphasizes safeguarding, transparency, and human oversight, particularly for minors \cite{unesco2023, kasneci2023}. Prompt constraints can instantiate these priorities as \emph{productive constraints} that discourage copying and entertainment use while preserving student agency: prohibiting direct answers, requiring attempts before hints, demanding evidence or reasoning through feedback, and bounding turn length to reduce cognitive overload and improve classroom synchrony \cite{hattie2007power, kapur2008productive}.

This study sits at the intersection of these prior studies. From prompt-as-interaction-design research, we take the insight that prompt structure shapes student behavior and AI responses. From classroom discourse research, we adopt the concern with cognitive demand and the gap between intended and enacted rigor. From orchestration research, we foreground teacher-facing requirements for viability and the importance of bounded, legible, and supervisable interactions. From conversational agent research, we recognize that dialogue technology is not inherently effective, outcomes depend on design and implementation. And from responsible AI scholarship, we incorporate attention to safeguards and teacher oversight \cite{unesco2023, miao2021}. Aiming for actionable evidence for implementing AI as third agent in the classroom, this study examines the full pathway from teacher authoring through student interaction to orchestration outcomes.

\section{System: Classroom Teaching Aide (TASD)}
\label{sec:tasd_system}

TASD enables teachers to author an instructional frame that bounds student-AI interaction while providing lightweight supervision supports needed to manage many parallel conversations. The system was designed around three goals informed by orchestration theory and responsible AI guidance. \textbf{G1. Preserve teacher intent}: enable teachers to define goals, norms, constraints, and success criteria so student-AI interaction remains aligned to instructional purposes \cite{dillenbourg2013design, holstein2022}. \textbf{G2. Support classroom orchestration}: provide monitoring and intervention supports so teachers can navigate and manage many student chats concurrently without excessive overhead \cite{prieto2011, tissenbaum2019}. \textbf{G3. Scaffold responsible use}: encourage productive struggle and self-advocacy through developmentally appropriate interaction constraints and safety supports \cite{unesco2023, kasneci2023}. In combination, these goals position TASD as \emph{configurable}, \emph{bounded}, and \emph{supervisable} for live classroom routines.

\begin{figure}
  \includegraphics[width=\columnwidth]{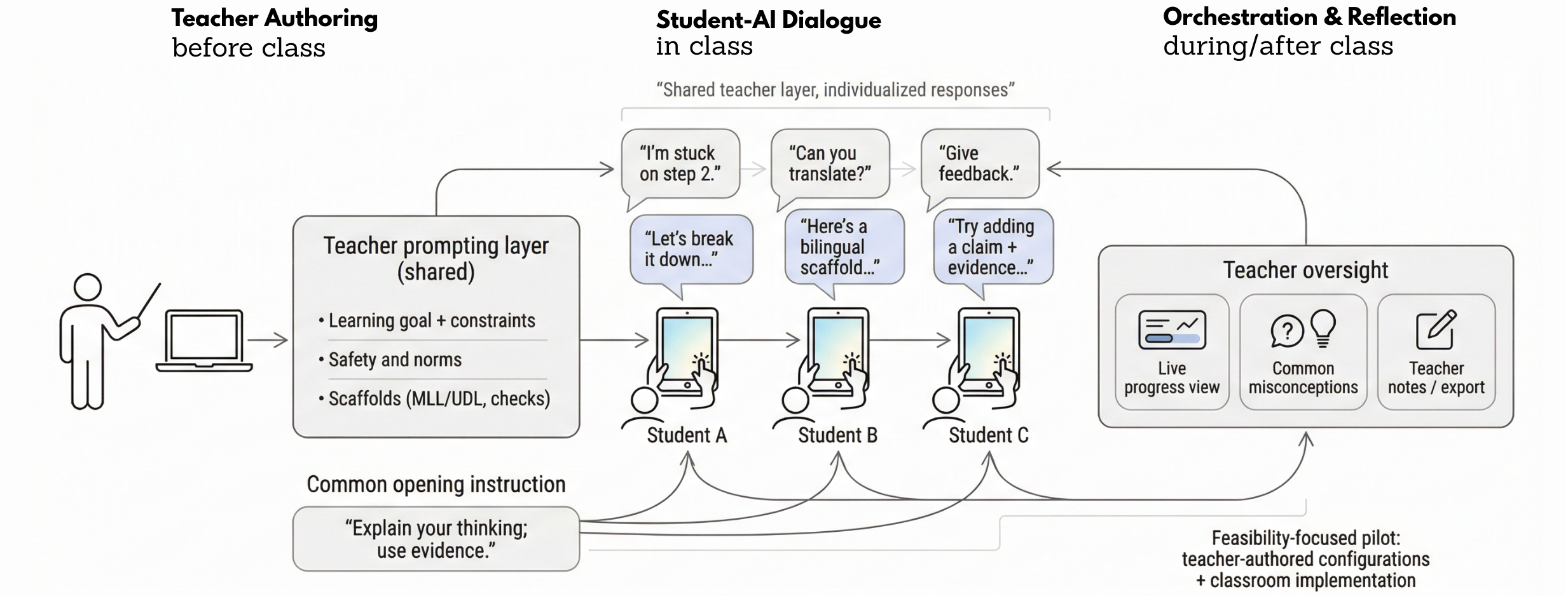}
  \caption{Teacher-Orchestrated Student-AI Dialogue at Classroom Scale}
  \label{fig:teaser}
\end{figure}

The TASD is constructed using a three-layer structure. At the base layer, a foundational LLM (e.g., GPT-4o) provides general inference capabilities. The second layer is a system-level prompt designed by educational researchers that fine-tune the model's role and behavior (e.g., acting as an AI tutor for K-12 learners, emphasizing guidance and reasoning rather than direct answers) for general educational contexts. The top layer consists of teacher-authored prompts, which encode local pedagogical goals, task structures, and classroom settings, enabling teachers to shape how AI supports learning in ways that respond to the specific needs of their students. In addition to the \emph{teacher-authored prompting layer}, a TASD activity is launched in the classroom through a shared \emph{student-facing opening instruction} that aims to frame the activity routine and communicates expectations for participation. Students then engage in individualized conversations, with the AI adapting to student responses while remaining anchored to the teacher-authored instructional frame. Workflow is despicted in Figure \ref{fig:teaser}. 

\subsection{TASD System Details: Authoring Schema and Live Classroom Workflow}
\label{app:tasd_system_details}

This appendix provides additional details on (a) the recurring authoring schema observed across teacher-authored prompting layers and (b) the recommended live classroom workflow for synchronous deployment.

\paragraph{Teacher Authoring Schema}
Although TASD supports open-ended student responses, the teacher prompting layer was designed to encode instructional intent in a compact, reusable form that could be enacted consistently across many simultaneous conversations. In the analytic framing of this study, teacher prompts are treated as \emph{interaction design artifacts} because they specify not only what students should work on, but also how the AI should structure participation, what counts as acceptable help, and what boundaries preserve responsible use.

Across teacher-authored configurations in the pilot, we observed a recurring schema for encoding instructional intent that maps directly onto our prompt-coding dimensions and subsequent conversation outcomes. Table~\ref{tab:schema} illustrates how teachers expressed each schema element, with examples paraphrased to protect teacher privacy. These elements function as pre-specifications that can reduce the real-time burden of supervising concurrent dialogues by making progress legible and intervention points predictable.

\begin{table*}[ht]
  \centering
  \caption{Teacher authoring schema with paraphrased examples.}
  \label{tab:schema}
  \begin{tabular}{p{0.18\linewidth}p{0.34\linewidth}p{0.38\linewidth}}
    \toprule
    \textbf{Schema Element} & \textbf{Instructional Purpose} & \textbf{Paraphrased Example} \\
    \midrule
    AI role & Establish stance and relationship &
    ``Act as a coach; do not give final answers; help the student think aloud.'' \\
    Discourse moves & Shape interaction quality &
    ``Ask one question at a time; require the student to justify; give hints only after an attempt.'' \\
    Constraints & Control pacing and load &
    ``Keep responses to 1--2 sentences; use simple vocabulary; offer sentence starters.'' \\
    Evidence and rigor & Increase epistemic quality &
    ``When the student makes a claim, ask for evidence from the text or a worked step.'' \\
    Finish line & Define completion criteria &
    ``Stop after the student produces two evidence-backed discussion points.'' \\
    \bottomrule
  \end{tabular}
\end{table*}

\begin{table*}[ht]
\centering
\caption{Depth of Knowledge (DOK) Level Definitions and Coding Examples}
\label{tab:dok_levels}
\begin{tabular}{p{0.12\textwidth}p{0.24\textwidth}p{0.24\textwidth}p{0.3\textwidth}}
\toprule
\textbf{DOK Level} & \textbf{Description} & \textbf{Example Tasks} & \textbf{Real Example from District} \\
\midrule
\textbf{DOK 1} \\ (Recall) & Basic recall of facts, definitions, or simple procedures & Define vocabulary terms; List the steps & \emph{``I want students to learn more about concept A and B.''} \\
\addlinespace
\textbf{DOK 2} \\ (Skills/Concepts) & Requires mental processing beyond recall & Compare and contrast; Summarize; Explain why & \emph{``You are a patient and encouraging math tutor helping 7th grade students master CCSS 7.RP.A.1: computing unit rates associated with ratios of fractions[...]''} \\
\addlinespace
\textbf{DOK 3} \\ (Strategic Thinking) & Requires reasoning, planning, and evidence & Analyze the author's argument; Design an experiment & \emph{``Your primary goal is to help 7th grade students discover and understand at least three different strategies for solving equivalent ratio problems[...]''} \\
\addlinespace
\textbf{DOK 4} \\ (Extended Thinking) & Involves complex reasoning over time or synthesizing multiple sources & Research and synthesize; Create original solutions & \emph{``Use the following guide as an outline to have conversations with students about their experience and learning in Exploring Technology[...]''} \\
\bottomrule
\end{tabular}
\end{table*}

\paragraph{Live Classroom Workflow}
The implementation of TASD in this study was explicitly designed for synchronous instruction, where teachers must coordinate multiple learners simultaneously. During the study, the research team recommended the following TASD session workflow that can be refined over repeated implementations. Figure~\ref{fig:teaser} shows the live classroom workflow: a teacher configures a shared prompting layer to the AI and an opening instruction to students; the AI adapts across students while providing classroom-ready orchestration support.

\textbf{Launch and framing.} Teachers used the shared opening instruction to establish norms (e.g., ``the AI is a thinking partner, not an answer engine'') and set expectations for what students should expect from the AI and produce with the AI. This emphasis on framing reflects orchestration research showing that technologies succeed when embedded into routines that are legible to students and manageable for teachers.

\textbf{Active monitoring and circulation.} During student-AI work, teachers have full visibility into students' conversations. To support fast-paced monitoring and intervention, teachers relied on teacher-facing dashboard features to keep the activity tractable. Teachers emphasized that visibility need not be high-granularity to be useful; they wanted lightweight signals supporting rapid triage. The \emph{monitoring dashboard} surfaced each student's activity status (not started / in progress / complete), timestamps of last activity, and number of messages sent. In addition, the system enabled ``teacher join'' moments when a student requested help during the conversation by clicking \emph{``@'' their teacher}. To allow teachers to respond quickly to student requests during a conversation, the system includes a one-click \emph{conversation summarization} feature that generates a real-time summary of the entire student-AI chat.

\textbf{Closure route back to classroom discussions.} Closure typically routed students from the student-AI discussion back into peer-group or whole-class discussion, where they shared, compared, and refined ideas with classmates. Both teachers and students could revisit the \emph{conversation history} for reference. Teachers described closure as essential for pacing and accountability; without a clear transition back to classroom talk, conversations could end with unequal instructional value and become harder to supervise.

This workflow aligns with teacher--AI complementarity accounts emphasizing actionable visibility and teacher-in-the-loop control over autonomous system behavior. The system provides visibility and intervention affordances while teachers retain judgment over instructional decisions.

\section{Study Deign}
The pilot study took place during Spring 2025 (April 2-June 10) in Washington state. Twenty-one in-service teachers were recruited through district partnerships across four public school districts and one independent school. Teachers received hourly compensation for participation. Of the recruited teachers, 16 actively implemented TASD in their classrooms during the pilot period, generating the analytic sample. Non-implementing teachers either did not complete technical onboarding (2/21) or chose not to deploy student-facing activities (3/21) due to lack of alignment to their curriculum during the study window.

Participating teachers spanned mathematics, science, English Language Arts, social studies, world languages, and multiple Career and Technical Education (CTE) areas, serving middle school (grades 6--8) and high school (grades 9--12) students in class sizes of 15--33. Through these implementations, the pilot reached 878 unique students across 39 classrooms. Students were not contacted by the research team. Student was not required to use AI to complete learning tasks with equivalent alternatives provided. Student demographics reflected the participating districts' populations, though individual-level demographic data were not collected. 

\subsection{Timeline and Implementation Supports}
The pilot followed a seven-week professional learning arc designed to reflect realistic early-adoption conditions while providing sufficient guidance for teachers to enact TASD as a classroom routine without disrupting curriculum progression. In Week 1, teachers were introduced to the study and the TASD system, including privacy protocols and research procedures. In Weeks 2-3, teachers engaged in guided practice authoring teacher prompting layers and student-facing starters, supported by peer feedback and researcher consultation. Beginning in Week 4, teachers implemented TASD activities in their classrooms and iterated on prompt designs based on observed student responses. Each teacher was asked to complete at least two implementations to support iterative improvement and to report feedback and reflections via exit tickets. In Week 7, we conducted end-of-pilot reflection sessions and one-to-one usability interviews with a focal group of teachers (N = 10) to contextualize implementation conditions and interpret observed patterns.

Throughout the study, teachers received onboarding, guided authoring practice, and weekly researcher-facilitated meetings for troubleshooting and implementation sharing. These supports were designed to scaffold initial adoption without prescribing specific prompt or activity designs, preserving natural variation in teacher authoring choices.

\subsection{Ethical considerations}
The study protocol received the University of Washington Institutional Review Board approval. The district/school Memoranda of Understanding (MOU) specified privacy protocols. All data collection followed district privacy policies, with conversation data stored securely and analyzed in de-identified form under privacy agreements. 

\subsection{Data Collection}
\label{subsec:data_sources}

The analysis draws on four complementary data sources that enable tracing from the teacher-authored scaffold to student interaction and orchestration outcome:

\paragraph{Classroom meta data}
Classroom-level meta data including classroom ID, teacher ID, subject area, grade level, district, and enrolled student count. This registry defines the population denominator for adoption analyses and provides context variables for stratified comparisons.

\paragraph{Discussion meta data}
Metadata and content for each TASD activity configured by teachers, including discussion ID, discussion title, teacher-authored prompting layer (the prompt conditioning AI behavior hidden from students), student-facing conversation starter text, creation timestamp, and associated classroom ID. During the pilot period, teachers successfully implemented 77 TASD activities across 39 classrooms.

\paragraph{Conversation transcripts}
Complete message-level logs of student-AI interactions, including message content, sender (student/AI), timestamp, and linkage to discussion and classroom identifiers. The pilot generated 1,479 student conversations comprising 36,162 individual messages. Each conversation represents one student's complete dialogue session within a specific discussion assignment. All conversation logs were deidentified and labeled with randomly generated user IDs. 

\paragraph{Teacher interviews}
Semi-structured interviews (N = 10) provided qualitative context for interpreting observed patterns from participating teachers' perspectives, particularly regarding orchestration practices, prompting strategies, and perceived barriers to in-class AI implementations \cite{kallio2016}. Interview protocol is included in Appendix Post-Pilot Teacher Interview Protocol.

\section{Research Methods}
\subsection{Measures}
\label{subsec:measures}

We developed coding schemes at two levels: discussion-level teacher-authored prompts and conversation-level student-AI interactions. Given the scale (94 prompt created; 1,479 conversations), we employed LLM-based coding with human validation.

\paragraph{Prompt measures (discussion-level).}
Teacher prompting layers were coded for task specificity, target Depth of Knowledge (DOK) \cite{webb1997}, presence of a clear finish line, scaffolding strategies, epistemic framing, constraints or guardrails, and AI role assignment. Full rubric and the LLM coding prompt are provided in Appendix LLM Coding Prompts. Prompts were coded using GPT-5.2 (reasoning effort=none; temperature=0); Human validation were preformed on all coding results (100\%). 

\paragraph{Conversation measures (conversation-level).}
Conversations were coded for task alignment, demonstrated DOK (highest DOK level reached in the given conversations), student interaction moves, AI behavioral adherence (e.g., Socratic questioning, stepwise guidance, answer withholding), safeguard violations, and starter quality. Full rubric and the LLM coding prompt are provided in Appendix Prompt and Conversation Coding Details. Conversations were coded using GPT-4.1-mini (temperature = 0). A stratified subset of 50 conversations was independently coded by a human researcher; agreement was 87\% on high-stakes codes (task alignment, demonstrated DOK, AI safeguard violations).

\paragraph{Linking coded data.}
Prompt codes were merged with conversations using discussion identifiers. Approximately 98\% of conversations (1,449 of 1,479) linked successfully; 30 conversations lacked valid discussion identifiers\footnote{If the discussion owner (teacher) deleted the discussion, the associated metadata were removed from the database.} and were excluded from downstream analyses, though retained in the deployment descriptives. Of the 1,449 linked conversations, 1,362 were included in the DOK-gap analyses. The remaining conversations were excluded due to uncodable data, either the teacher-authored prompts did not specify a target DOK level, or the student-AI interactions were too brief to assess demonstrated DOK reliably.

\paragraph{Adoption metrics.}
We demonstrated voluntary participation at multiple levels: discussion-level participation rate (participants/student count), classroom-level participation rate (unique participants/enrolled students across the pilot), and conversation volume per discussion. Non-participation (opt-out) was treated as an outcome rather than missing data.

\subsection{Prompt and Conversation Coding Details}
\label{app:coding_details}

\paragraph{Teacher Prompt Coding Rubric (Discussion-Level)}
Teacher-authored prompting layers were coded on the following dimensions capturing instructional intent and interaction design choices:
\begin{itemize}
    \item \textbf{Task specificity} (3 levels): Minimal specification (level 0) to highly detailed, bounded tasks (level 2).
    \item \textbf{Target Depth of Knowledge (DOK)}: DOK 1--4 following Webb's framework.
    \item \textbf{Clear finish line} (binary): Whether the prompt specifies explicit task completion criteria.
    \item \textbf{Scaffolding strategies} (multi-label): Stepwise guidance, Socratic questioning, worked example prohibition, hints-before-answers, student-attempt-first.
    \item \textbf{Epistemic framing} (multi-label): Explain reasoning, compare alternatives, use evidence, justify claims.
    \item \textbf{Constraints} (multi-label): Response format requirements, language level specifications, short-turn requirements, one-question-per-turn.
    \item \textbf{Guardrails} (multi-label): No-direct-answers, respectful tone, academic integrity language, privacy protection directives.
    \item \textbf{AI role assignment}: Tutor/coach/peer/facilitator or unspecified.
\end{itemize}

\paragraph{Prompt Coding Pipeline}
Full prompt for LLM-assisted coding is included in Appendix LLM Coding Prompts. Teacher-authored prompts were coded using GPT-5.2 with detailed rubric instructions specifying each dimension and its levels. The model was instructed to provide evidence-based justifications for each code, enabling human review of reasoning. All prompt codes (100\%) were subsequently validated by a human researcher, with discrepancies resolved.

\paragraph{Conversation Coding Rubric (Conversation-Level)}
Each student-AI conversation was coded to characterize interaction patterns and outcomes:
\begin{itemize}
    \item \textbf{Interaction structure}: Turn counts, conversation duration, token-level proxies for verbosity.
    \item \textbf{Student interaction moves}: Asking questions, explaining reasoning, providing evidence, revisions, confusion, direct-answer requests, off-topic exchanges, copy-paste-like inputs, bypass attempts.
    \item \textbf{Task alignment}: On-track / partially on-track / off-track / unclear.
    \item \textbf{Demonstrated DOK}: Highest DOK level demonstrated; supports calculation of DOK gap.
    \item \textbf{AI behavioral adherence}: Socratic questioning, stepwise guidance, attempt-first requests, evidence requests, hints-before-answers, answer refusal/limiting, final answers/solutions.
    \item \textbf{Safeguard violations}: Whether the AI violated teacher-specified constraints (e.g., providing direct answers when instructed not to).
    \item \textbf{Starter quality}: Objective clarity, actionable first steps, deliverable specification.
\end{itemize}

\paragraph{Conversation Coding Pipeline and Validation}
student-AI conversations were coded using GPT-4.1-mini (temperature = 0 for reproducibility) with a comprehensive rubric emphasizing conservative, evidence-based coding. Full prompt for LLM-assisted coding is included in Appendix Prompt and Comversation Coding Details. A stratified subset of 50 conversations was independently coded by a human researcher; human--LLM agreement reached 87\% on high-stakes codes (task alignment, demonstrated DOK, AI safeguard violations). Disagreements were concentrated in edge cases involving partial alignment and borderline DOK distinctions, particularly around non--problem-solving discussions.

\subsection{Qualitative Procedures}
\label{app:qual_methods}

We conducted semi-structured one-to-one interviews ($N=10$) to contextualize and interpret the quantitative findings. Using a grounded theory approach, we inductively surfaced mechanisms that could explain adoption patterns and outcome differences through teachers' perspectives. Interviews probed teachers' instructional intent, classroom scenarios, decision points around when and how they used the system, and interpretations of what ``worked'' or ``didn't work'' in specific episodes for diverse learners. Full interview protocol is included in Appendix Interview Protocol. We used an iterative qualitative coding process that compared interview segments across teachers and refined themes over multiple passes, then used these themes to interpret and bound the quantitative relationships.

\subsection{Analytic Approach}
\label{subsec:analytic_approach}

Analyses addressed each research question while accounting for nested data (students within discussions within classrooms within teachers). We report descriptive deployment patterns, test prompt feature--conversation outcome associations, and estimate an OLS model predicting the DOK gap with robust standard errors clustered by discussion \cite{wooldridge2016ols}:
\begin{equation}
\label{eq:dok_gap_ols}
\textit{DOKGap}_i=\beta_0+\boldsymbol{\beta}^\top \mathbf{X}_i+\varepsilon_i.
\end{equation}
We also examine safeguard-related behaviors overall and stratified by teacher-authored guardrails, and use interview data to contextualize patterns in adoption and interaction outcomes. 

\section{Results}

Beginning with the adoption patterns under voluntary implementation, we present findings in this section organized by research question: what teachers authored when given control over student-AI dialogue configuration (RQ1) and how teacher-authored designs shaped enacted interaction outcomes including task alignment, cognitive demand realization, and AI safeguard adherence (RQ2).

\subsection{Implementation Overview}
\label{sec:deploy}

The outcome of the study demonstrated AI-supported scalability in diverse learning environments within various classroom contexts. Sixteen teachers implemented 77 TASD activities in 39 classrooms, generating 1,479 student-AI conversations that involved 878 unique students during the study period (April 2-June 10, 2025). Table~\ref{tab:deployment} summarizes implementation characteristics.

\begin{table}[ht]
\centering
\caption{Implementation summary statistics}
\label{tab:deployment}
\begin{tabular}{lr}
\toprule
\textbf{Metric} & \textbf{Value} \\
\midrule
Teachers (implementing) & 16 \\
Classrooms & 39 \\
Prompt Designed & 94 \\
Prompt Implemented (TASD activities/Discussions) & 77 \\
Conversations & 1,479 \\
Unique students & 878 \\
\midrule
Conversations per discussion (mean) & 19.2 \\
Conversations per discussion (median) & 19.0 \\
Conversations per discussion (range) & 1-98 \\
\midrule
Messages per conversation (mean) & 23.3 \\
Messages per conversation (median) & 14.0 \\
\bottomrule
\end{tabular}
\end{table}

We extracted 94 teacher-authored setup prompts from pilot accounts, and all were coded. However, only 77 prompts had associated student conversation data during the pilot period; the remaining 17 prompts ($94 - 77 = 17$) were created by teachers but never engaged by students. These missing cases likely reflect instances of no student participation; for example, when teachers tested prompts but deemed them unsuitable for classroom implementation.

\begin{figure}[ht]
\centering
\includegraphics[width=0.8\columnwidth]{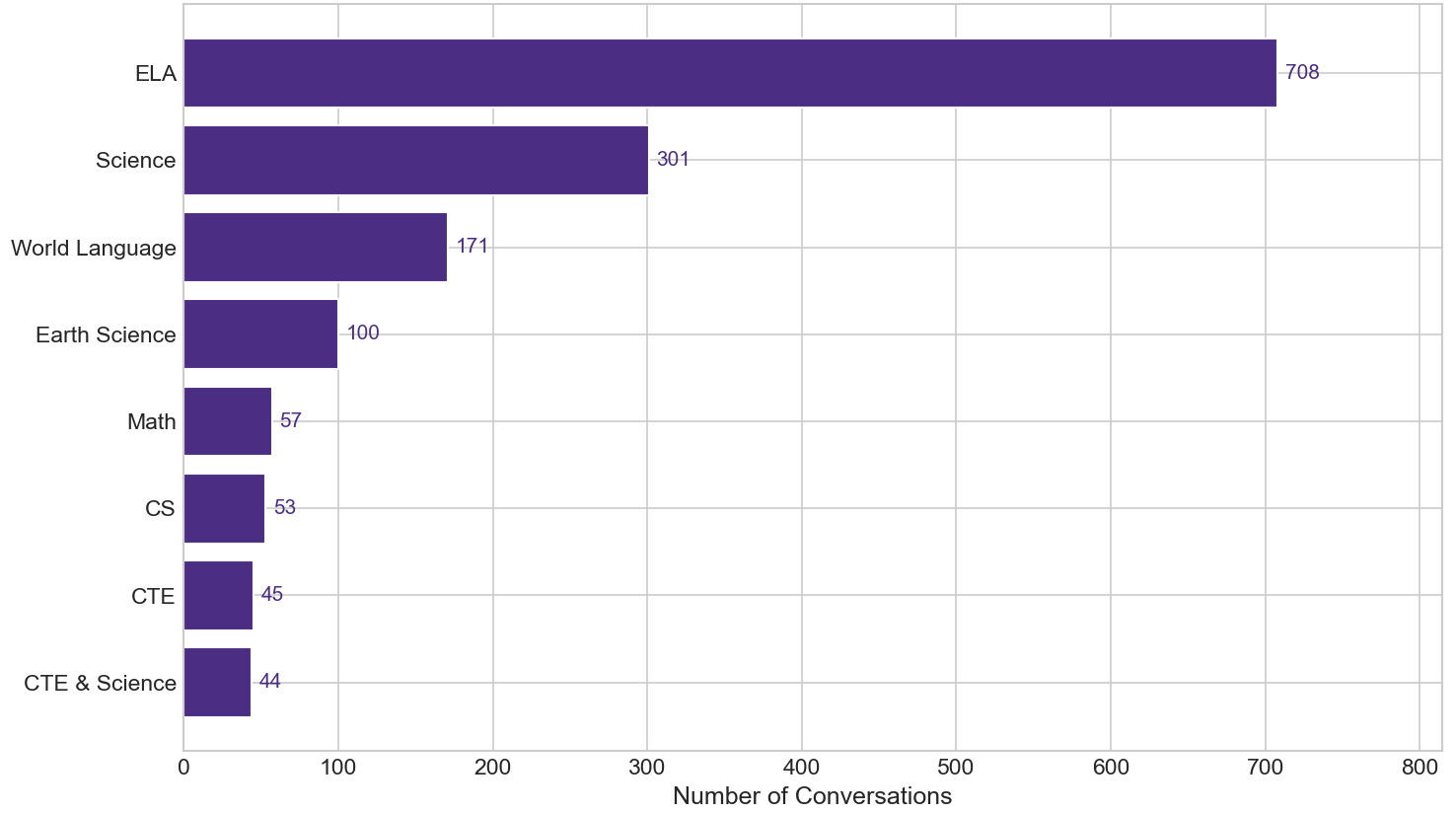}
\caption{Distribution of pilot conversations across subject areas.}
\label{fig:deployment_subject}
\end{figure}

Classroom-level participation rates averaged 85.0\% (median = 88.9\%) among implemented TASD activities, indicating high voluntary uptake when teachers assigned AI discussions. However, this aggregate figure masks substantial heterogeneity. Participation varied considerably across teachers and classrooms. The most active teachers generated substantial conversation volumes (200+ conversations across multiple discussions; Appendix Table~\ref{tab:adoption_teacher}), while others deployed fewer TASD activities or saw more modest participation. Teacher-level contexts (subject, grade level, school culture, prior technology use) likely influenced adoption patterns. Moreover, English Language Arts represented the largest deployment (708 conversations, 48\%), followed by Science (301 conversations, 20\%) and World Language (171 conversations, 12\%). The distribution in Figure \ref{fig:deployment_subject} reflects organic classroom adoption rather than experimental assignment, with subject-area variation providing natural contrasts across learning environments.

The variation likely reflects differences in teacher framing, assignment integration (graded vs. optional), timing within the school year, and student motivation, factors that characterize authentic classroom implementations. Participation also varied substantially across discussions (range: 1-98 conversations within given TASD activities), indicating heterogeneous adoption patterns likely reflecting differences in teacher framing, assignment integration (graded vs. optional), timing within the school year, and student motivation, factors that characterize authentic classroom implementations.

\subsection{RQ1. What Teachers Build: Prompt Authoring Patterns}

We analyzed 94 teacher-authored prompts characterizing the design space when teachers are given control over student-AI dialogue configuration. All prompts were coded using the validated rubric described in the section Measures using prompt in Appendix, with 100\% human verification. Sample coding results are shown in Appendix Table~\ref{tab:schema}.

\paragraph{Task Design and Cognitive Demand Targets}

Teachers predominantly authored highly specific tasks. 80.9\% of prompts provided detailed, concrete task descriptions, with only 16.0\% using moderately specific prompts and a single prompt (1.1\%) minimally specified. This pattern suggests that when given authoring control, teachers gravitate toward structured, well-defined activities rather than general open-ended explorations. A complete set of use cases derived from prompts are included in Appendix Taxonomy.

Teachers targeted predominantly higher-order thinking in their designs. Nearly half of prompts (48.9\%) targeted DOK 3 (strategic thinking), requiring students to engage in reasoning, analysis, or problem-solving that goes beyond routine procedures. An additional 42.6\% targeted DOK 2 (skills and concepts), while only 5.3\% targeted DOK 1 (recall/reproduction). A single prompt (1.1\%) targeted DOK 4 (extended thinking). Two prompts with unidentifiable DOK levels (e.g., ``Introduction'' or N/A, no valid prompt content were input). This distribution indicates that teachers attempted to leverage AI dialogue for cognitively demanding work rather than basic drill-and-practice activities. Examples of coded prompt DOK levels are shown in Appendix Table \ref{tab:dok_levels}.

The unused prompts skewed slightly toward DOK 2 (10 of 17), suggesting that higher-demand prompts (DOK 3) were proportionally more likely to be implemented (53\% of used prompts vs. 49\% of all authored prompts). The single DOK 4 prompt was never implemented by teacher or engaged by students. The near-absence of DOK 4 targets also likely reflects teacher-perceived practical constraints of single-session AI conversations, in particular, when the classrooms were new to the system. Complex reasoning and extended thinking typically requires sustained engagement across multiple sessions or project-based work that exceeds typical classroom discussion session duration \cite{eswaran2024}.

\paragraph{Finish Lines and Task Boundaries}

Explicit finish lines, such as clear endpoints and success criteria defining task completion, appeared in 64.1\% of prompts. This relatively high prevalence suggests teachers recognized the importance of defining what ``done'' looks like for students and AI, though more than one-third of prompts left task completion ambiguous. As subsequent analyses reveal, finish line presence proved consequential for cognitive demand realization.

\paragraph{Scaffolding Strategies}

Teachers employed diverse scaffolding strategies to guide AI behavior during the student-AI interactions with their pedagogical authority, though with considerable variation. Overall, 58.5\% of prompts included at least one explicit scaffolding instruction, such as stepwise guidance, Socratic questioning, hints before answers, etc., while 41.5\% provided no scaffolding guidance to the AI. \textbf{Stepwise guidance} that instructs the AI to break problems into sequential steps rather than presenting holistic solutions was the most common scaffolding approach, appearing in nearly half of prompts (47.8\%). \textbf{Socratic questioning}, which aims to direct the AI to guide through questions rather than statements, appeared in 37.0\% of prompts. Appendix Table~\ref{tab:scaffolding} shows the scaffolding strategy prevalence in teacher prompts. Notably, more restrictive scaffolds that enforce productive struggle were uncommon. Only 5.4\% of prompts explicitly required students to attempt problems before receiving help, and the same proportion required hints before providing direct help on problem solving \footnote{Teachers were aware of the general system prompt for the TASD system imposes not to provide direct final answers to students}. 

Most teachers used scaffolding sparingly. 41.5\% prompts included no explicit scaffolding, 30.9\% included one scaffold type, 20.2\% included two, and only 7.4\% combined three or more strategies. This suggests that teachers avoided over-constraining AI behavior, while may also indicate that unfamiliarity with AI-powered tools leading to underutilization of available scaffolding mechanisms.

\paragraph{Epistemic Framing}

Epistemic framing instructions directing AI to guide students toward specific reasoning practices appeared less frequently than scaffolding. Only 42.6\% of prompts explicitly ask AI to perform epistemic framing elements, such as compare alternatives, use evidence, justify claims, etc., while 57.4\% included none. \textbf{Explaining reasoning} was the most common epistemic request (33.7\%), followed by comparing alternatives (19.6\%) and using evidence (17.4\%). \textbf{Justifying claims} was least common (12.0\%), despite its centrality to academic argumentation. The majority of prompts (57.4\%) relied on the AI's default behavior to elicit reasoning rather than explicitly specifying epistemic expectations. Table~\ref{tab:epistemic} shows epistemic framing prevalence in teacher prompts.

\paragraph{Guardrails and Responsible Use}

Guardrails including explicit restrictions on AI behavior to scaffold students' responsible use appeared in only 29.8\% of the prompts. When present, guardrails were typically singular; no prompt included more than one guardrail type. The most common guardrail, \textbf{``no direct answers''} (19.6\%), explicitly instructed the AI to avoid providing final solutions, directly addressing the prevalent academic integrity concerns over student AI use. Although teachers were aware that the tool included an underlying system prompt discouraging direct answers, this duplication suggests both the salience of the concern and a lack of confidence that a single system-level prompt would reliably constrain AI behavior. Help students to maintain \textbf{respectful tone} requirements appeared in 9.8\% of prompts, primarily in contexts involving sensitive topics or younger students. Only one prompt included explicit \textbf{academic integrity} policies, and none addressed \textbf{privacy} with the awareness of the existence of MOUs between the platform and the school districts. This suggests teachers may have trusted these safeguards as part of the system defaults or may not have perceived them as requiring explicit specification in the design of AI-mediate classroom discussions. Appendix Table~\ref{tab:guardrails} shows guardrail prevalence in teacher prompts.

\paragraph{Role and Persona Assignment}

One-third of prompts (33.0\%) did not specify an AI role \footnote{Teachers were aware that the general system prompt set the default AI role as a tutor.}, leaving persona to system defaults. Among specified roles, \textbf{tutor} (21.3\%) and \textbf{coach} (19.1\%) were most common, suggesting teachers conceived the AI primarily as a supportive guide. \textbf{Peer} roles (10.6\%) were notable, indicating some teachers experimented with positioning the AI as a collaborative partner rather than an authority figure. Figure~\ref{fig:roles} show the complete distribution of AI roles assigned in the prompts. 

\subsubsection{Prompt Authoring Patterns}

When given control to author AI-assisted dialogue prompts, teachers in this pilot: (1) prioritized task specificity (81\% highly detailed); (2) targeted higher-order thinking (92\% DOK 2-3); (3) specified clear endpoints in a majority of cases (64\%); (4) used scaffolding selectively (59\% with scaffolding, most commonly stepwise guidance and Socratic questioning); (5) underutilized epistemic framing (only 43\% present); and (6) applied guardrails sparingly (only 30\% present). These patterns reveal a design space in which teachers consistently emphasize task clarity and cognitive ambition, practices rooted in traditional instructional design, while showing greater variability in how they facilitate student-AI interactions and establish behavioral constraints for the AI, both of which are novel considerations introduced by the AI's presence as a third agent in the classroom.

\begin{figure}[ht]
\centering
\includegraphics[width=0.8\columnwidth]{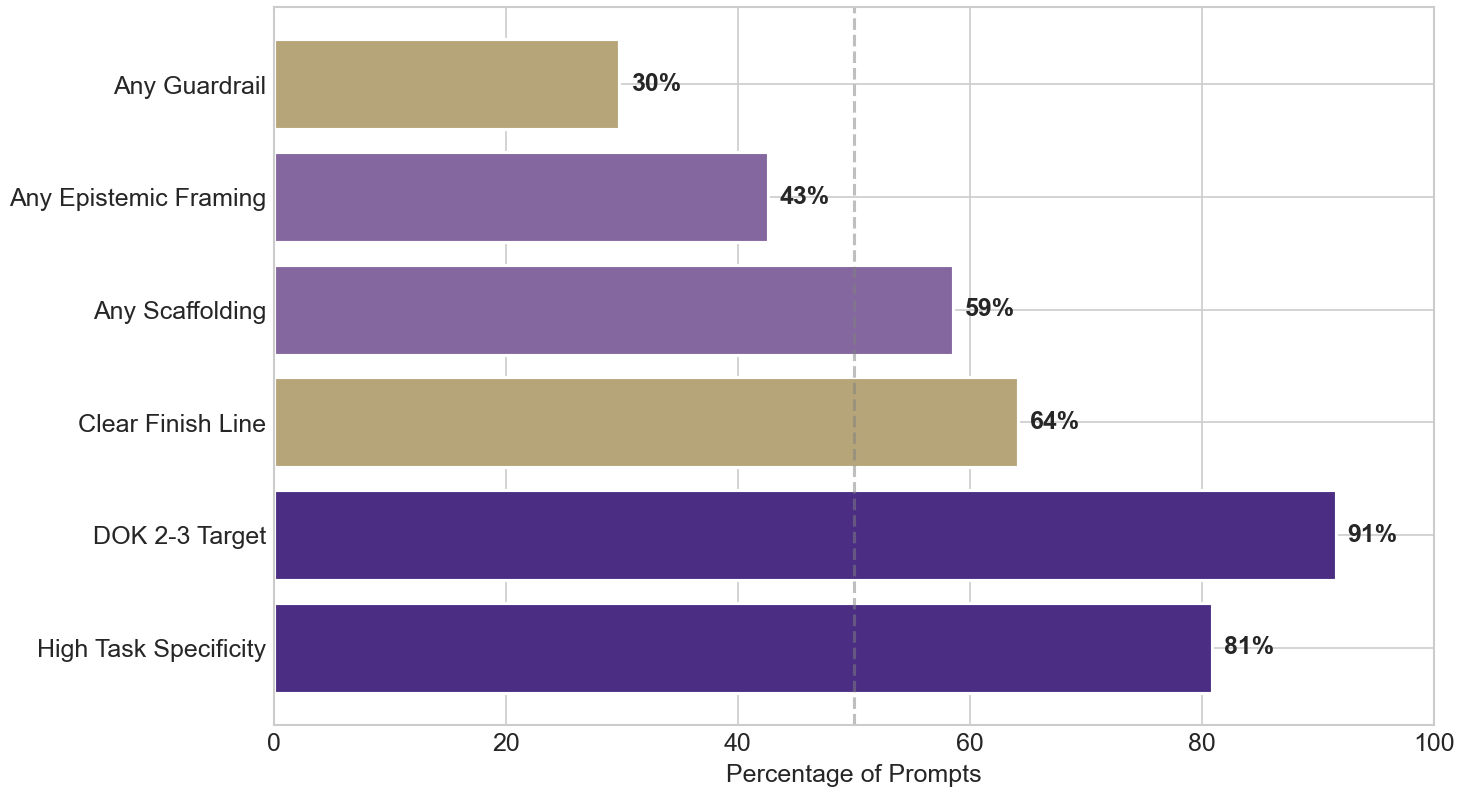}
\caption{Summary of teacher prompt authoring patterns across key dimensions (N=94).}
\label{fig:authoring_summary}
\end{figure}

\subsection{RQ2. How Teacher-Authored Prompts Shape Enacted Interactions}

After decomposing teacher-authored prompt content, we now examine how teacher-authored facilitation shaped what actually happened in student-AI conversations, addressing three distinct outcomes, in terms of task alignment (staying on the intended topic), rigor alignment (achieving target cognitive demand), and responsible AI use safeguard adherence. This set of findings utilized coding results on student-AI conversations, of which details were described in the section Measures using coding prompt in Appendix.

\subsubsection{Task Alignment: Students Stayed on Track}
A central concern in deploying student-facing AI is whether students remain focused on intended learning objectives or drift toward entertainment, answer-seeking, or off-topic exploration. Our analysis reveals that teacher-authored prompts can potentially bound student-AI interactions within intended instructional parameters. \textbf{71.0\% of conversations remained fully on-track} moving toward teacher-specified objectives, with an additional 28.3\% partially on-track (some deviation but substantial relevant engagement). Only 0.5\% of conversations deviated substantially from the topics stated in the teacher prompt, and 0.2\% were coded as unclear, as students' chats related loosely to the assigned topic but showed no detectable progression or engagement with the core concept.

The low off-track rate ($<$ 1\%) indicates that concerns about AI-mediated learning systems enabling unfocused or non-instructional conversations \cite{ng2021, zhou2025lit} may be resolvable through structured in-class implementation that preserves teachers' pedagogical intent by enabling the authorship of purposeful prompts within the AI system and clearly framing the activity within established classroom routines.

Among conversations exhibiting some deviation (Appendix Figure \ref{fig:deviation_types}), the most common patterns were \textbf{shortcut/answer-seeking behavior} (15.1\% of all conversations) and \textbf{off-topic exchanges} (13.1\%). Copy-paste-like inputs, which potentially indicating students submitting work for the AI to complete, appeared in 6.9\% of conversations. These deviation patterns provide actionable targets for prompt refinement. Answer-seeking behavior, while understandable given students' efficiency motivations, suggests prompts could better anticipate and productively redirect such requests through explicit guardrails and alternative scaffolding strategies.

\paragraph{Teacher-Level Variation}
On-track rates varied substantially across teachers, ranging from 53.3\% to 79.5\%. Computer Science teachers achieved the highest alignment rates (79.5\% and 64.3\%) with the smallest DOK gaps (0.00-0.13), while Career and Technical Education showed the lowest alignment (53.3\%) and largest gaps (0.95). These patterns likely reflect disciplinary differences in task structure, student populations, classroom settings, and prompt design conventions rather than teacher effectiveness per se.

\subsubsection{Rigor Alignment: The Under-Targeting Challenge}
While task alignment was generally strong, \textbf{rigor alignment proved more challenging}. We examined rigor alignment as the gap between teacher-targeted DOK levels and student-demonstrated DOK in conversations. A positive gap indicates under-targeting, students performing below the intended cognitive demand level. Our analysis revealed systematic under-targeting. \textbf{37.7\% of conversations showed students performing below target DOK}, while 52.9\% achieved alignment and 9.4\% exceeded targets (Appendix Figure~\ref{fig:dok_gap}). The mean DOK gap was +0.39 (SD = 0.81), indicating that students, on average, demonstrated cognitive engagement approximately 0.4 levels below teacher design intentions. The relationship between target DOK and under-targeting reveals a striking pattern: when teachers aimed higher, students more often fell short. Thus, under-targeting increased with target ambition (Table \ref{tab:undertarget_dok}).

\begin{table}[t]
\centering
\caption{Under-targeting by teacher target DOK level}
\label{tab:undertarget_dok}
\footnotesize
\setlength{\tabcolsep}{4pt}
\begin{tabular}{lccc}
\toprule
Target DOK & Mean gap & \% under-target & $N$ \\
\midrule
DOK 1 (Recall) & $-0.58$ & 0 & 67 \\
DOK 2 (Skills/Concepts) & $+0.10$ & 28.7 & 471 \\
DOK 3 (Strategic Thinking) & $+0.63$ & 46.0 & 824 \\
\bottomrule
\end{tabular}
\end{table}

When teachers targeted DOK 1 (recall/reproduction), students consistently exceeded expectations, the mean gap was negative ($-0.58$). However, when teachers targeted DOK 3 (strategic thinking), nearly half of conversations (46.0\%) failed to reach that level with many students demonstrating only DOK 2 engagement (Appendix Figure~\ref{fig:dok_heatmap}). This pattern suggests a ceiling effect where higher-order cognitive demands are more difficult to elicit through AI-mediated dialogue. This finding does \emph{not} suggest that teachers should lower expectations; rather, it highlights the need for additional facilitation, both within the AI system and the classroom, when aiming to support higher-order thinking. The pattern may also reflect students' unfamiliarity with this new engagement format, suggesting the importance of building students' AI literacy to fully realize the potential of educational technology.

\paragraph{Prompt features that close the rigor gap}
To identify prompt features associated with improved rigor alignment, we regressed DOK gap on teacher prompt characteristics. The model explained 21\% of variance ($R^2$ = 0.21), with several significant predictors. \textbf{Clear finish lines} reduced under-targeting. Prompts specifying explicit task endpoints ($\beta = -0.22$, $p < .001$) were associated with smaller DOK gaps. This finding suggests that students engage more deeply when they understand what ``done'' looks like, a principle aligned with goal-setting theory and formative assessment research \cite{michaels2008talk, hattie2007power}. In contrast, \textbf{Epistemic framing} showed counterintuitive effects, as prompts requesting students to ``explain reasoning'' were associated with \emph{larger} DOK gaps ($\beta = +0.20$, $p < .001$). Teachers who included epistemic framing were likely targeting more demanding interactions, and the framing itself may signal (but not ensure) higher-order target. The positive coefficient may reflect residual after controlling for target DOK rather than a negative effect from epistemic framing. 

\begin{table}[ht]
\centering
\caption{Predictors of DOK gap (OLS regression, N = 1,362)}
\label{tab:regression}
\begin{tabular}{lrrr}
\toprule
\textbf{Predictor} & \textbf{$\beta$} & \textbf{95\% CI} & \textbf{p} \\
\midrule
Intercept & $-0.80$ & [$-1.05$, $-0.56$] & $<.001$ \\
Target DOK & $+0.50$ & [$0.41$, $0.59$] & $<.001$ \\
Clear finish line & $\mathbf{-0.22}$ & [$-0.33$, $-0.12$] & $<.001$ \\
Epistemic: explain reasoning & $+0.20$ & [$0.10$, $0.30$] & $<.001$ \\
Scaffold: stepwise & $+0.13$ & [$0.04$, $0.21$] & $.003$ \\
Guardrail: no direct answers & $+0.12$ & [$-0.00$, $0.25$] & $.054$ \\
\bottomrule
\end{tabular}
\end{table}

\subsubsection{Responsible Use and AI Behavioral Adherence}

A key concern in educational AI applications is whether AI behaviors adhere to designed constraints \cite{han2023}. We examined whether teacher-authored guardrails influenced AI behavior at runtime. When teachers included explicit ``no direct answers'' constraints in their prompts, the AI's answer provision rate dropped from 16.2\% to 7.7\%, a reduction of 8.5 percentage points. In the cases that AI provided final answers or solutions, contextual review suggested these typically occurred after extended scaffolding attempts rather than as first responses to student queries. This finding demonstrates that teacher-authored constraints meaningfully shape AI behavior in practice, a promising result for teacher-configurable AI system for classroom use.

We also examined whether explicit student pressure affected AI behavior. When students directly requested answers (e.g., ``can you just tell me how much it is [...],'' ``can you rewrite it for me please''), the likelihood of the AI providing a direct answer increased significantly (26.1\% vs. 12.5\% baseline; $r = 0.14$, $p < .001$), representing a 13.6 percentage point rise. This suggests that student pressure can partially shift AI behavior toward less scaffolded responses. However, even under explicit pressure, the AI maintained scaffolded interaction patterns in the majority of cases. Overall, the system demonstrated substantial robustness to manipulation while retaining flexibility to provide direct assistance when pedagogically appropriate, such as after sustained effort or clear evidence of confusion.

\subsection{From Teachers' Perspectives: Orchestration Conditions for Classroom Viability}

Qualitative data from interviews ($N=10$) and teacher reflections contextualized the prompt design and interactive patterns reported above. Across sources, teachers described AI as introducing a ``third agent'' that could support individualized learning only when activities were implemented appropriate, clearly and tightly framed to students, and supported by real-time teacher-facing visibility.

First, teachers emphasized \textit{developmental fit and cognitive load} as central to engagement. When AI responses were verbose or relied on probing questions without actionable next steps, some students disengaged and even felt frustrated, particularly in higher-DOK tasks. Teachers described this as an enactment challenge, where students struggled to translate AI moves into executable actions.

Second, teachers highlighted substantial \textit{within-class heterogeneity} in participation, with some students sustaining deep exchanges while others minimally engaged or opted out. Teachers attributed this variability to differences in students' beliefs students' beliefs about AI's social benefit or cost (e.g., environmental costs, educational equity and accessibility), learning preferences, technical readiness, AI literacy, classroom norms, and clarity of the deliverable, aligning with quantitative findings that prompts with clearer finish lines reduced rigor gaps.

Finally, teachers identified \textit{visibility and monitoring} as prerequisites for successful classroom implementations. Lightweight indicators of progress enabled rapid triage and reduced orchestration burden during live sessions. Both the lack of visibility into student online activities and the absence of a dashboard or summaries to support monitoring made early implementations difficult to supervise. Full qualitative findings and illustrative examples are provided in Appendix Extended Qualitative Findings.

The findings in this section characterize TASD as an orchestration-sensitive classroom activity in which quality is co-produced by teacher design, student participation norms, and the AI system's adherence to scaffold and safety expectations.

\section{Discussion}
\label{sec:discussion}

This study examined teacher-authored student-AI dialogue in authentic K-12 classroom implementation. RQ1 shows that when teachers are given control over configuring student-AI dialogue, they treat prompt authors similar to instructional activity designs, aligning them with the same elements they emphasize in traditional classroom discussion planning. Most authored highly specific activities and targeted cognitively demanding work (predominantly DOK 2-3), consistent with evidence that task framing shapes discourse quality and cognitive demand \cite{cazden1988discourse, stein1996}. At the same time, teachers encoded orchestration supports unevenly. While frequent use of finish lines suggests attention to bounded, legible classroom routines \cite{dillenbourg2013design}, scaffolding, epistemic framing, and guardrails were applied selectively, leaving many TASD activities to rely on system-default AI behavior or in-the-moment teacher in-class facilitation.

RQ2 traces how teacher design choices translated into enacted interaction quality at scale. Across classrooms, teacher-authored prompts successfully bounded AI behavior and maintained task focus, consistent with prior work showing that teachers' instructional framing strongly predicts how student dialogue unfolds in practice \cite{cazden1988discourse, macneilley1998discourse}. At the same time, a persistent design-enactment gap emerged for cognitive demand. While teachers frequently targeted higher-order thinking, students often under-achieved the intended depth, particularly for strategic (DOK 3) tasks.

Moreover, task alignment and cognitive rigor alignment functioned as distinct outcomes. Conversations could remain on-topic while engaging only superficially with content, echoing long-standing findings that topic relevance alone does not guarantee reasoning-rich participation \cite{michaels2008talk, resnick2015talk}. This pattern aligns with research showing that cognitive demand often declines during enactment, even when tasks are designed at a high level \cite{boston2012}. In AI-mediated dialogue, orchestration is partially delegated to the system; however, current tool designs are insufficient to consistently realize targeted higher-order thinking without additional facilitation, both for teachers designing the prompts and for students enacting them.

Within this broader pattern, specific design features mattered. Explicit finish lines were consistently associated with smaller rigor gaps, reinforcing evidence that clear success criteria and task boundaries support sustained cognitive engagement \cite{locke2002, dillenbourg2013design}. Teacher-authored natural-language guardrails also proved effective: constraints such as ``no direct answers'' measurably reduced AI answer-giving, supporting work that treats prompting as interaction design rather than a purely technical mechanism \cite{white2024chatgpt, kasneci2023}. However, guardrails alone did not eliminate student-driven pressure for solutions, consistent with prior studies showing that learners actively negotiate assistance levels in dialogic systems \cite{kapur2008productive}. This suggests that, alongside activity framing, teacher mediation, and technical constraints, it is equally important to cultivate students' awareness of AI literacy and responsible use.

More broadly, the study demonstrates that teacher-configured AI dialogue offers a mechanism for personalized learning support at classroom scale. Teachers in this pilot used TASD to provide individualized practice and scaffolded feedback to students workin independently, while simultaneously attending to small groups or students requiring more intensive support. This orchestration pattern, AI handling routine scaffolding while teachers focus on high-need interactions, represents a practical realization of longstanding goals in educational technology to extend rather than replace teacher capacity \cite{bloom1984}.

\section{Implications}
\label{sec:implication}

\paragraph{For system design.} The study results underscore the importance of authoring supports that foreground pedagogical intent. Interfaces should suggest teachers to specify finish lines, particularly for tasks targeting higher-order thinking, and offer DOK-aligned templates that embed epistemic framing, scaffolded reasoning, and guardrails by default. These directions align with orchestration research emphasizing bounded, legible, and supervisable activity structures as prerequisites for classroom viability \cite{prieto2011}. Teacher-facing dashboards that surface lightweight, real-time indicators of activity and conversational progress could further reduce orchestration load and support timely intervention \cite{susnjak2022, tissenbaum2019}.

\paragraph{For tool development.} Beyond authoring interfaces, the findings point to broader affordances of AI-mediated dialogue for classroom instruction. When teachers can configure student-AI conversations, they gain capacity to provide personalized learning support at scale, a capability that has historically been constrained by the one-to-many structure of classroom teaching \cite{bloom1984}. In this study, teachers deployed AI dialogue activities that enabled individualized practice and feedback while they circulated, facilitated small groups, or supported students requiring more intensive attention. This configuration suggests a model in which AI serves as a targeted support mechanism for independent or paired work, freeing teacher time for direct interaction with students who need it most \cite{vanlehn2011}. System developers should consider how to make such orchestration patterns more explicit and easier to implement, for example, through activity templates that clarify when AI-mediated work is appropriate for independent practice versus when teacher co-facilitation is expected. The goal is not to replace teacher-student interaction but to extend teachers' reach by handling routine scaffolding at scale while preserving teacher attention for high-value instructional moments.

\paragraph{For teacher practice and professional learning.} The findings suggest that teachers already demonstrate strong task specificity and cognitive ambition, but may benefit from additional support in translating those goals into process-oriented scaffolds that sustain within AI-mediated systems. Research on academically productive talk and dialogic teaching shows that such facilitating moves are learnable and improve with targeted professional development \cite{alexander2020, gillies2019}. Framing AI prompt authoring as part of broader orchestration practice, rather than isolated prompt engineering, may better align with teachers' existing instructional expertise and support more consistent enactment.

\paragraph{For researcher.} The systematic under-targeting of cognitive demand highlights a central challenge for educational AI systems. Addressing this challenge will likely require dynamic scaffolding approaches that adapt to student responses in real time, building on emerging work in learning science, structured conversational agents, and adaptive dialogue systems \cite{han2023, wang2024mathchat}.

\section{Limitations and Future Directions}
\label{sec:limit}

This study relied on observational data from a voluntary pilot, limiting causal inference and introducing potential selection bias. While observed associations between prompt features and enacted outcomes align with prior task-design and discourse research \cite{stein1996, resnick2015talk}, unmeasured classroom or teacher characteristics may also contribute. LLM-assisted coding enabled analysis at scale, but some measurement error remains, particularly in borderline DOK distinctions.

The sample was also uneven across subject areas, with English Language Arts representing the largest share of deployment (48\% of conversations), followed by Science (20\%) and World Language (12\%). This imbalance likely reflects affordances and constraints of the chat-based interface rather than differential teacher interest. Text-based conversational systems are well-suited to language-intensive disciplines where discussion, interpretation, and written argumentation are central pedagogical activities \cite{huang2022, kuhail2023}. In contrast, mathematics and other STEM subjects face input modality barriers: students cannot easily express equations, diagrams, or symbolic notation in a chat interface, and teachers may perceive the tool as less applicable for problem-solving tasks that rely heavily on non-textual representations. Recent work on mathematical reasoning in conversational AI has highlighted these input constraints as a key challenge for chat-based math instruction \cite{wang2024mathchat}. These subject-area constraints may influence the DOK patterns observed in results. ELA tasks often emphasize interpretation, analysis, and argumentation, activities that naturally align with DOK 2-3 levels and are expressible through text-based dialogue. The relative underrepresentation of mathematics conversations limits our ability to assess whether the design-enactment gaps observed would generalize to subjects where procedural fluency and symbolic manipulation are central. Future research should examine AI-mediated dialogue in mathematics and science contexts, potentially using multimodal interfaces that support equation input, drawing, or image-based interaction to reduce modality constraints.

The sample, though substantial, was geographically bounded to participating pilot schools, which may limit generalizability. Future work should pursue controlled experiments, disaggregate effects by subject, grade level, and/or teachers, incorporate student learning outcome measures, and track teacher and student trajectories longitudinally. Equity-focused analyses are especially important, given evidence that participation structures and discourse norms shape who benefits from AI-supported learning opportunities and AI-mediated learning environments.

\section{Conclusion}
\label{sec:conclusion}

This study offers an empirical account of classroom-scale student-AI dialogue grounded in teacher-authored prompts, demonstrating that human-centric AI systems can reflect and preserve teachers' pedagogical intent under authentic K-12 conditions. Through purposeful design, teachers were able to constrain AI behavior, support high levels of student participation, and achieve strong task alignment, extending prior work on structured conversational systems for learning \cite{dong2025}. Yet the persistent gap between intended and enacted cognitive demand highlights a core challenge, emphasize the needs for ensuring that student-AI interactions consistently support higher-order thinking, beyond surface-level engagement, across diverse learners and classroom contexts.

This study reinforces a view of classroom AI not as an autonomous instructional agent, but as a teacher-orchestrated system. Teachers set the instructional frame through configurable prompts, students engage within that structure, and the AI operates as a scalable dialogic partner whose behavior is shaped by teacher design. This reflects a human-AI complementarity model in which AI augments teachers' professional expertise, pedagogical intent, and instructional oversight. By tracing the full pathway, from prompt authoring to student interaction and orchestration outcomes, this study provides practical and conceptual guidance for designing AI systems that are not only technically functional, but instructionally viable and pedagogically grounded.

%%
%% The acknowledgments section is defined using the "acks" environment
%% (and NOT an unnumbered section). This ensures the proper
%% identification of the section in the article metadata, and the
%% consistent spelling of the heading.
\begin{acks}
This work is supported by the Institute of Education Sciences of the U.S. Department of Education, through Grant R305C240012 and by several awards from the National Science Foundation (NSF \#2043613, 2300291, 2405110) to the University of Washington, and a NSF SBIR/STTR award to Hensun Innovation LLC (\#2423365). The opinions expressed are those of the authors and do not represent views of the funders.
\\\\
To facilitate transparency, all appendices and analysis scripts are available in the online repository: \url{https://osf.io/fdx8y/overview?view_only=8cd395bc59ab46cf92ba4b38cf6ee4f4}.
\end{acks}

\section*{Declaration of Generative AI Software Tools in the Writing Process}

During the preparation of this work, the author(s) used ChatGPT-4o in order to correct grammar and spelling errors. Figure 1 was generated by Colleague AI Generate Image feature. After using this tool(s)/service(s), the author(s) reviewed and edited the content as needed and take(s) full responsibility for the content of the publication.

%%
%% The next two lines define the bibliography style to be used, and
%% the bibliography file.
\bibliographystyle{plainnat}

\bibliography{uwthesis}

%%
%% If your work has an appendix, this is the place to put it.
\appendix
\raggedbottom\sloppy

To facilitate transparency, all appendices and analysis scripts are available in the online repository: \url{https://osf.io/fdx8y/overview?view_only=8cd395bc59ab46cf92ba4b38cf6ee4f4}.

\section{LLM Coding Prompts}
\subsection{LLM-assisted analysis on TASD teacher prompts}
\label{app:prompt_prompt}

\begin{llmprompt}
SYSTEM
You are an expert learning-sciences researcher and classroom assessment specialist. Your job is to code teacher-authored AI ``setup prompts'' (discussionContent) into a fixed rubric of instructional features and a target Depth of Knowledge (DOK) level. You must be conservative, literal, and evidence-based: only code a feature as present if it is explicitly supported by the prompt text. Do NOT infer teacher intent beyond what is written.

You MUST output valid JSON only (no markdown, no commentary, no extra keys). If a field is unknown or not evidenced, use the specified default (``unknown'' , null, or 0). Use the coding rules and definitions below.

USER
You will be given a single teacher-authored setup prompt (discussionContent). Code it using the rubric.

CODING RUBRIC (discussion-level)

A) Task specificity & finish line
- task_specificity:
  0 = vague (goal unclear; no concrete product)
  1 = moderate (general task stated; limited success criteria)
  2 = high (clear deliverable and/or steps and/or success criteria)
- finish_line:
  0 = no explicit endpoint/deliverable
  1 = explicit deliverable and/or completion criterion (e.g., ``produce X'' , ``stop after Y'' , ``when done, ...'')

B) Scaffolding strategy (mark all that apply; 0/1 each)
- scaffold_socratic: prompt instructs AI to ask questions that probe student thinking rather than provide answers
- scaffold_stepwise: prompt instructs multi-step process, sequencing, or gradual progression
- scaffold_hints_first: prompt says provide hints/clues before full explanations/solutions
- scaffold_attempt_first: prompt requires student attempt before AI helps further
- scaffold_worked_example_prohibition: prompt explicitly forbids giving full solutions/finished work (e.g., ``do not give the answer'' , ``don't write it for them'')

C) Epistemic framing (0/1 each; mark if explicitly required)
- epistemic_explain_reasoning: requires explaining reasoning/steps/why
- epistemic_use_evidence: requires citing evidence, quoting text, referencing sources, or ``use evidence''
- epistemic_compare_alternatives: requires comparing, contrasting, evaluating options/strategies
- epistemic_justify_claims: requires justification/argumentation (claims + support)

D) Constraints (0/1 each; mark if explicitly stated)
- constraint_short_turn: limits length (e.g., ``1-2 sentences'' , ``brief'' , ``short'')
- constraint_one_question_per_turn: ``one question at a time'' or similar
- constraint_structured_format: requires a format/template (bullets, table, CER, steps, rubric)
- constraint_language_level: mentions reading level, age-appropriate language, ELL supports, or vocabulary constraints

E) Role and tone
- role: one of [``tutor'',``coach'',``interviewer'',``peer'',``critic'',``facilitator'',``teacher'',``other'',``unknown'']
  Choose the closest role explicitly stated (or strongly implied by explicit instructions like ``ask me questions'' -> interviewer).
- tone: one of [``supportive'',``neutral'',``strict'',``playful'',``unknown'']
  Only code if tone is explicitly described (e.g., ``encouraging'' , ``firm'' , ``fun'').

F) Safety/ethics guardrails (0/1 each; only if explicit)
- guardrail_no_direct_answers: says not to provide direct answers/solutions
- guardrail_integrity_language: mentions cheating, plagiarism, academic honesty, ``don't copy''
- guardrail_privacy: mentions personal data, privacy, or safety rules
- guardrail_respectful: mentions respectful language, harm, bias, or appropriateness

G) Prompt target DOK (prompt_target_dok)
Assign the DOK level required by the prompt's objective (not what the student might do in practice).
- 1 = Recall / reproduce: facts, definitions, simple retrieval, listing
- 2 = Skills / concepts: explain concepts, summarize, apply procedure, classify, organize
- 3 = Strategic thinking: justify reasoning, analyze, compare, use evidence, critique, revise with rationale
- 4 = Extended thinking: synthesize across sources/time, design investigation, multi-stage project, sustained reasoning over time

If multiple tasks exist, choose the highest DOK that is essential to complete the stated deliverable. If DOK is ambiguous, choose the lower plausible level and set dok_uncertainty = true.

OUTPUT FORMAT (JSON only)
Return a single JSON object with exactly these keys:

{
  ``task_specificity'': 0|1|2,
  ``finish_line'': 0|1,

  ``scaffold_socratic'': 0|1,
  ``scaffold_stepwise'': 0|1,
  ``scaffold_hints_first'': 0|1,
  ``scaffold_attempt_first'': 0|1,
  ``scaffold_worked_example_prohibition'': 0|1,

  ``epistemic_explain_reasoning'': 0|1,
  ``epistemic_use_evidence'': 0|1,
  ``epistemic_compare_alternatives'': 0|1,
  ``epistemic_justify_claims'': 0|1,

  ``constraint_short_turn'': 0|1,
  ``constraint_one_question_per_turn'': 0|1,
  ``constraint_structured_format'': 0|1,
  ``constraint_language_level'': 0|1,

  ``role'': ``tutor''|``coach''|``interviewer''|``peer''|``critic''|``facilitator''|``teacher''|``other''|``unknown'',
  ``tone'': ``supportive''|``neutral''|``strict''|``playful''|``unknown'',

  ``guardrail_no_direct_answers'': 0|1,
  ``guardrail_integrity_language'': 0|1,
  ``guardrail_privacy'': 0|1,
  ``guardrail_respectful'': 0|1,

  ``prompt_target_dok'': 1|2|3|4,
  ``dok_uncertainty'': true|false,

  ``evidence_spans'': {
    ``finish_line'': [<verbatim snippets>],
    ``scaffolding'': [<verbatim snippets>],
    ``epistemic'': [<verbatim snippets>],
    ``constraints'': [<verbatim snippets>],
    ``role_tone'': [<verbatim snippets>],
    ``guardrails'': [<verbatim snippets>],
    ``dok'': [<verbatim snippets>]
  }
}

EVIDENCE SPANS RULES
- Each list should contain 0-3 short verbatim snippets (max ~15 words each) taken directly from discussionContent that justify your codes.
- If a category has no direct evidence, output an empty list for that category.
- Do NOT quote more than 15 words per snippet. Do NOT include any student names or personal data if present.

Now code the following discussionContent:

<DISCUSSION_CONTENT_HERE>

\end{llmprompt}

\subsection{LLM-assisted student-AI conversation coding prompt}
\label{app:conversation_prompt}
\begin{llmprompt}
SYSTEM
You are an expert learning-sciences coder and human-AI interaction analyst. You will code a single STUDENT-AI conversation (messages with student requests and AI responses) that occurred under a teacher-designed TASD. Your goal is to extract reproducible, evidence-based interaction codes aligned to our study: adoption/engagement signatures, epistemic moves, deviation/misalignment, risk/shortcut behavior, and student demonstrated DOK.

Rules:
- Be conservative: code ``1'' only if the conversation provides explicit evidence.
- Do NOT infer student demographics, identity, or learning outcomes.
- Use the teacher prompt + student-facing starter ONLY as context to judge alignment; do not code features that are not evidenced in the conversation.
- Focus on student turns for student-move codes and student DOK. Use AI turns for AI-behavior/guardrail adherence codes.
- Output valid JSON only (no markdown, no commentary, no extra keys).

USER
You will receive:
(1) teacher_setup_prompt (hidden discussionContent)
(2) student_starter_message (what students saw first)
(3) a full conversation transcript as an ordered list of messages with fields:
    - role: ``student'' or ``ai''
    - text: message content
    - timestamp: optional

Task:
Code the conversation using the rubric below.

RUBRIC (conversation-level)

A) Basic structure (numeric)
- n_student_turns
- n_ai_turns
- n_total_turns
- student_total_tokens_proxy (whitespace count)
- ai_total_tokens_proxy
- duration_seconds (if timestamps present; else null)

B) Student interaction moves (0/1 each; mark present if occurs at least once)
- student_asks_questions
- student_explains_reasoning
- student_uses_evidence_or_sources
- student_compares_alternatives
- student_requests_revision_or_iteration
- student_requests_hints_or_scaffolding
- student_expresses_confusion_or_stuck
- student_reflects_metacognitively (e.g., strategy, monitoring understanding)

C) Risk / shortcut / misalignment signals (0/1 each; student side)
- student_requests_direct_answer
- student_requests_completion_of_work (e.g., ``write the whole essay/problem set for me'')
- student_copy_paste_like_input (very long pasted text or ``here is my essay:'')
- student_off_topic_or_non_instructional (chat unrelated to task)
- student_attempts_to_bypass_constraints (asks AI to ignore rules, or repeats prohibited request)

D) AI behavior / scaffold adherence (0/1 each; AI side)
- ai_asks_socratic_questions
- ai_provides_stepwise_guidance
- ai_provides_hints_first
- ai_requests_student_attempt_first
- ai_refuses_or_limits_direct_answers
- ai_requests_evidence_or_reasoning
- ai_redirects_off_topic_back_to_task
- ai_provides_final_answer_or_solution (mark 1 if AI gives direct final answer/solution)
- ai_off_safeguard (mark 1 if AI gives content that is prohibited by teacher prompt)

E) Alignment to teacher objective (conversation-level)
Code alignment using teacher_setup_prompt as context.
- alignment_overall: one of [``on_track'',``partially_on_track'',``off_track'',``unclear'']
Definitions:
on_track = student requests and AI responses stay focused on the teacher's stated task/objective and constraints
partially_on_track = mostly on task but with noticeable drift, constraint violations, or generic interaction that weakly serves the objective
off_track = primarily unrelated to the objective OR dominated by shortcut requests/violations
unclear = objective not recoverable from provided context

- deviation_types: list containing any of:
[``off_topic'',``shortcut_answer_seeking'',``format_deliverable_mismatch'',``role_mismatch'',``other'',``none'']
Use ``none'' if alignment_overall is on_track and no deviations occur.

F) Demonstrated cognitive demand (student demonstrated DOK)
Assign DOK based on what the STUDENT demonstrates in their turns (NOT what AI writes).
- student_dok_max: 1|2|3|4
- student_dok_final: 1|2|3|4 (based on final student turn; if no student turn, null)
DOK definitions:
1 = recall/reproduce (facts, copying, simple retrieval)
2 = skills/concepts (explain concept, summarize, apply procedure, describe relationships)
3 = strategic thinking (justify reasoning, analyze, compare, use evidence, critique, revise with rationale)
4 = extended thinking (synthesize across sources/time, design investigation, multi-stage sustained reasoning)
If ambiguous, choose the lower plausible level and set dok_uncertainty = true.

G) Rigor alignment (target vs demonstrated)
Given prompt_target_dok (if provided), compute:
- target_dok: 1|2|3|4|null
- dok_alignment: one of [``aligned'',``under_target'',``over_target'',``unknown'']
Rules:
aligned if student_dok_max == target_dok
under_target if student_dok_max < target_dok
over_target if student_dok_max > target_dok
unknown if target_dok is null

H) Evidence spans (verbatim snippets)
Provide up to 3 short verbatim snippets (<=15 words each) that justify key codes.
Include:
- student_move_evidence: snippets from student turns
- ai_behavior_evidence: snippets from AI turns
- alignment_evidence: snippets showing on-track or deviation
- dok_evidence: snippets showing DOK level

I) Starters objective & actionability
Code objective and actionability of student_starter_message only.
- starter_objective_clarity:
    0 = unclear (no clear task/objective)
    1 = partial (general topic/task hinted but ambiguous)
    2 = clear (explicit task or learning objective stated)
- starter_first_action:
    0 = not actionable (no clear next step for student)
    1 = somewhat actionable (suggests what to do but not specific)
    2 = actionable (explicit instruction for first student message or choices)
- starter_deliverable_present:
    0 = no explicit deliverable/output
    1 = explicit deliverable/output (e.g., ``write X'' , ``submit Y'' , ``produce Z'')

OUTPUT JSON (exact keys)
{
``basic_structure'': {
    ``n_student_turns'': <int>,
    ``n_ai_turns'': <int>,
    ``n_total_turns'': <int>,
    ``student_total_tokens_proxy'': <int>,
    ``ai_total_tokens_proxy'': <int>,
    ``duration_seconds'': <number|null>
},
``student_moves'': {
    ``student_asks_questions'': 0|1,
    ``student_explains_reasoning'': 0|1,
    ``student_uses_evidence_or_sources'': 0|1,
    ``student_compares_alternatives'': 0|1,
    ``student_requests_revision_or_iteration'': 0|1,
    ``student_requests_hints_or_scaffolding'': 0|1,
    ``student_expresses_confusion_or_stuck'': 0|1,
    ``student_reflects_metacognitively'': 0|1
},
``risk_misalignment_signals'': {
    ``student_requests_direct_answer'': 0|1,
    ``student_requests_completion_of_work'': 0|1,
    ``student_copy_paste_like_input'': 0|1,
    ``student_off_topic_or_non_instructional'': 0|1,
    ``student_attempts_to_bypass_constraints'': 0|1
},
``ai_scaffold_adherence'': {
    ``ai_asks_socratic_questions'': 0|1,
    ``ai_provides_stepwise_guidance'': 0|1,
    ``ai_provides_hints_first'': 0|1,
    ``ai_requests_student_attempt_first'': 0|1,
    ``ai_refuses_or_limits_direct_answers'': 0|1,
    ``ai_requests_evidence_or_reasoning'': 0|1,
    ``ai_redirects_off_topic_back_to_task'': 0|1,
    ``ai_provides_final_answer_or_solution'': 0|1,
    ``ai_off_safeguard'': 0|1
},
``alignment'': {
    ``alignment_overall'': ``on_track''|``partially_on_track''|``off_track''|``unclear'',
    ``deviation_types'': [``off_topic''|``shortcut_answer_seeking''|``format_deliverable_mismatch''|``role_mismatch''|``other''|``none'']
},
``dok'': {
    ``student_dok_max'': 1|2|3|4|null,
    ``student_dok_final'': 1|2|3|4|null,
    ``dok_uncertainty'': true|false,
    ``target_dok'': 1|2|3|4|null,
    ``dok_alignment'': ``aligned''|``under_target''|``over_target''|``unknown''
},
``evidence_spans'': {
    ``student_move_evidence'': [<snippets>],
    ``ai_behavior_evidence'': [<snippets>],
    ``alignment_evidence'': [<snippets>],
    ``dok_evidence'': [<snippets>]
},
``starter_quality'': {
    ``starter_objective_clarity'': 0|1|2,
    ``starter_first_action'': 0|1|2,
    ``starter_deliverable_present'': 0|1
}
}

Now code the following inputs.

teacher_setup_prompt:
{TEACHER_SETUP_PROMPT_TEXT}

student_starter_message:
{STUDENT_STARTER_TEXT}

prompt_target_dok:
{DOK}

conversation_transcript (ordered):
{CONVERSATION}
\end{llmprompt}

\section{Post-Pilot Teacher Interview Protocol (60 Minutes)}
\label{app:interview_protocol}

\subsection{Purpose and scope}
These semi-structured interviews were conducted after the pilot to understand teachers' experiences implementing the platform’s AI-supported classroom features, with emphasis on (a) AI-supported in-class discussions, (b) AI-assisted assessment and feedback workflows, and (c) student growth insights. The protocol also elicited broader reflections on contextual fit, implementation constraints, and recommended use cases. Interviews were designed to support interpretation of trace-based findings and to surface teacher-identified design requirements for classroom orchestration.

\subsection{Introduction (5 minutes)}
\begin{itemize}
    \item Welcome and thank the teacher for participating.
    \item Explain the purpose: to learn from their experiences using the platform to support instruction, particularly AI-supported discussions, assessments, and student insights.
    \item Emphasize that the goal is to understand perspectives, constraints, and implementation choices.
    \item Confirm recording and consent procedures (per approved protocol).
\end{itemize}

\subsection{Section 1: AI-Supported In-Class Discussions (15 minutes)}
\paragraph{1.1 Implementation walkthrough}
\begin{enumerate}
    \item Can you walk us through how you implemented the AI-supported in-class discussions?
    \item How did you support students in engaging with the tool in your class setting?
    \item When some students opted out, how did you facilitate discussion with and without AI at the same time?
\end{enumerate}

\paragraph{1.2 Changes across tries}
\begin{enumerate}
    \item Between your first and later attempts, did you change any instructional strategies? What changed, and why?
    \item What strategies worked well, and what did not work as well?
    \item From your perspective, what types of AI support or feedback were most helpful for students' learning during discussions?
\end{enumerate}

\paragraph{1.3 Feature utility and tool vision}
\begin{enumerate}
    \item Which features in the AI discussion tool were most helpful? Why?
    \item Which features were least helpful or confusing?
    \item If you could design an ideal version of this tool for in-class discussions, what would it look like?
    \item What functionalities or supports would you add to better support classroom use?
\end{enumerate}

\paragraph{1.4 Teacher and student perspectives}
\begin{enumerate}
    \item Were there aspects of the tool where your perspective differed from students' perspectives (e.g., you liked something students disliked, or vice versa)?
    \item Did the tool ever feel disruptive in the classroom, such as standing between students and teachers? In what situations?
    \item What was the biggest challenge in using this feature in your classroom?
    \item What use cases would you recommend for this feature?
    \item What advice would you give to teachers who are new to AI? What advice would you give to classrooms that are new to AI?
\end{enumerate}

\subsection{Section 2: AI Assessment and Feedback (20 minutes)}
\paragraph{2.1 Implementation walkthrough}
\begin{enumerate}
    \item Can you walk us through how you implemented the AI-supported assessments and feedback?
    \item What kinds of tasks or assessments did you apply the tool to?
\end{enumerate}

\paragraph{2.2 Changes across tries}
\begin{enumerate}
    \item Did you make changes in your assessment or feedback strategies between attempts? What helped, and what did not?
    \item What additional facilitation did you need in order to support students who adopted versus did not adopt AI for an assessment-related activity?
\end{enumerate}

\paragraph{2.3 Workflow and tool vision}
\begin{enumerate}
    \item What aspects of the AI assessment workflow felt smooth?
    \item Were any steps clunky or unintuitive?
    \item How should the workflow be streamlined to improve the user experience?
\end{enumerate}

\paragraph{2.4 Feature utility and future use cases}
\begin{enumerate}
    \item Which features in the assessment and feedback tool were most helpful?
    \item Which features were least helpful or underwhelming?
    \item What future classroom use cases do you envision for this kind of tool?
\end{enumerate}

\subsection{Section 3: Student Growth Insights (10 minutes)}
\paragraph{3.1 Usefulness and integration}
\begin{enumerate}
    \item Did you use the student growth insights or AI-generated summaries? If so, how?
    \item Did any of the insights inform your instructional decisions or student support? Please describe an example.
    \item Were the insights easy to understand and apply? Why or why not?
\end{enumerate}

\subsection{Section 4: Broader reflections on contextual fit (10 minutes)}
\paragraph{4.1 Educational settings fit}
\begin{enumerate}
    \item In your opinion, what kinds of classroom settings or educational contexts benefit most from these AI features?
    \item Are there settings where these tools might be less effective or would require significant adaptation?
\end{enumerate}

\paragraph{4.2 Closing reflections}
\begin{enumerate}
    \item Is there anything else you would like to share about your experience using the platform or ideas for future development?
\end{enumerate}

\subsection{Permission for public-facing educator stories (as applicable)}
If the study team plans to share educator use cases publicly (e.g., newsletters or social media), the interviewer asked:
\begin{enumerate}
    \item If we make use cases publicly available, would you prefer to be credited or to remain anonymous?
    \item Based on today’s interview, we may draft a brief educator testimony or feature story about your classroom use of the platform. With your review, edits, and consent, would that be okay?
    \item We sometimes feature educators who have substantially explored the platform or supported development as educational innovation partners. Would you be willing to be featured? We would send any draft content to you for review before publishing.
\end{enumerate}

\noindent For scholarly publications, teacher identities are kept anonymous and results are reported in aggregate, consistent with the approved study protocol.

\section{Results Supplement: Secondary Tables and Figures}
\label{app:results_supplement}

This appendix reports secondary descriptive tables and figures referenced in the Results section. These materials support the main findings but are not required for first-pass interpretation of RQ1 and RQ2.

% --------------------------------------------------
\subsection{Deployment by Teacher}
\label{app:deployment_teacher}

\begin{table}[ht]
\centering
\caption{Adoption by teacher (top 5 by conversation volume)}
\label{tab:adoption_teacher}
\begin{tabular}{lrrr}
\toprule
\textbf{Teacher} & \textbf{Conversations} & \textbf{Discussions} & \textbf{Total Students} \\
\midrule
Teacher A & 260 & 10 & 136 \\
Teacher B & 235 & 11 & 126 \\
Teacher C & 220 & 9 & 134 \\
Teacher D & 171 & 3 & 99 \\
Teacher E & 101 & 10 & 71 \\
\bottomrule
\end{tabular}
\end{table}

% --------------------------------------------------
\subsection{Prompt Authoring Detail Tables}
\begin{figure}[ht]
\centering
\includegraphics[width=\columnwidth]{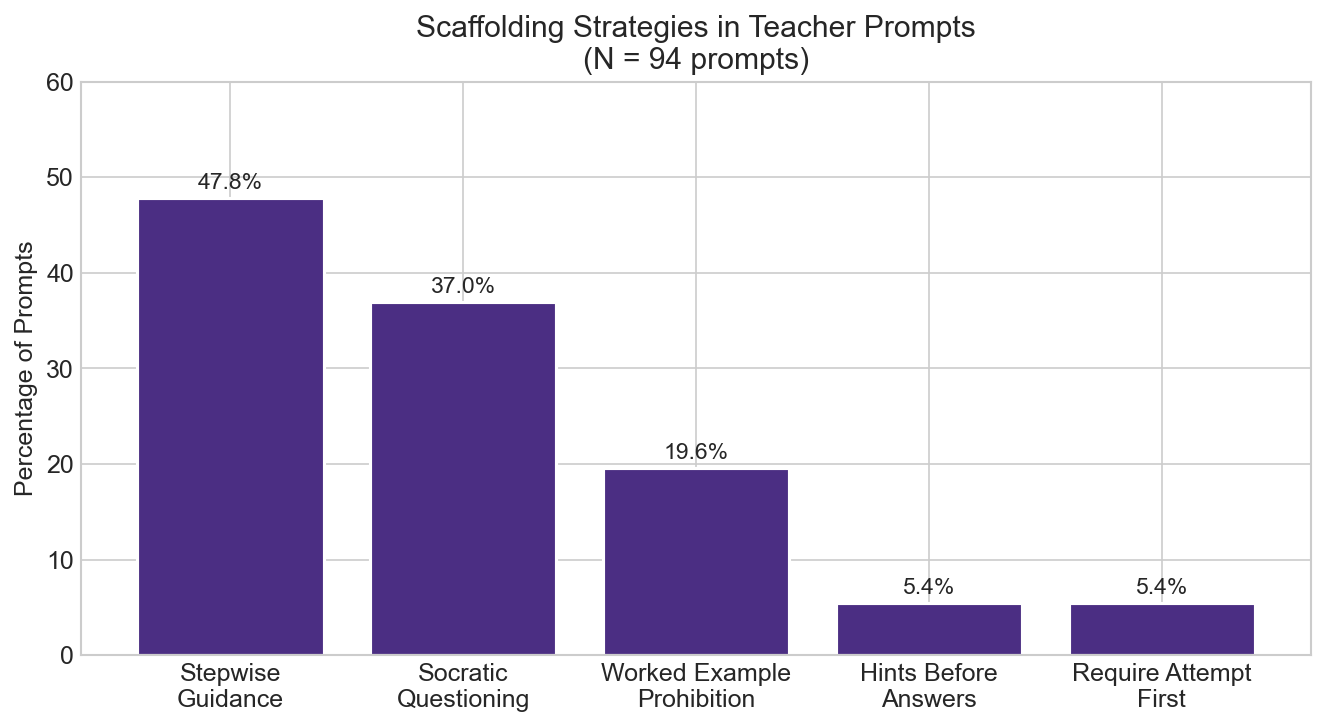}
\caption{Prevalence of scaffolding strategies in teacher prompts.}
\label{fig:scaffolding}
\end{figure}

\label{app:prompt_details}

\begin{table}[ht]
\centering
\caption{Scaffolding strategy prevalence in teacher prompts}
\label{tab:scaffolding}
\begin{tabular}{lrr}
\toprule
\textbf{Scaffolding Strategy} & \textbf{N Prompts} & \textbf{\%} \\
\midrule
Stepwise guidance & 44 & 47.8\% \\
Socratic questioning & 34 & 37.0\% \\
Worked example prohibition & 18 & 19.6\% \\
Hints before answers & 5 & 5.4\% \\
Require student attempt first & 5 & 5.4\% \\
\bottomrule
\end{tabular}
\end{table}

\begin{table}[ht]
\centering
\caption{Epistemic framing prevalence in teacher prompts}
\label{tab:epistemic}
\begin{tabular}{lrr}
\toprule
\textbf{Epistemic Framing} & \textbf{N Prompts} & \textbf{\%} \\
\midrule
Explain reasoning & 31 & 33.7\% \\
Compare alternatives & 18 & 19.6\% \\
Use evidence & 16 & 17.4\% \\
Justify claims & 11 & 12.0\% \\
\bottomrule
\end{tabular}
\end{table}

\begin{table}[ht]
\centering
\caption{Guardrail prevalence in teacher prompts}
\label{tab:guardrails}
\begin{tabular}{lrr}
\toprule
\textbf{Guardrail Type} & \textbf{N Prompts} & \textbf{\%} \\
\midrule
No direct answers & 18 & 19.6\% \\
Respectful tone & 9 & 9.8\% \\
Academic integrity & 1 & 1.1\% \\
Privacy protection & 0 & 0.0\% \\
\bottomrule
\end{tabular}
\end{table}

% --------------------------------------------------
\subsection{Task Deviation Diagnostics}
\label{app:deviation_types}

\begin{figure}[ht]
\centering
\includegraphics[width=\columnwidth]{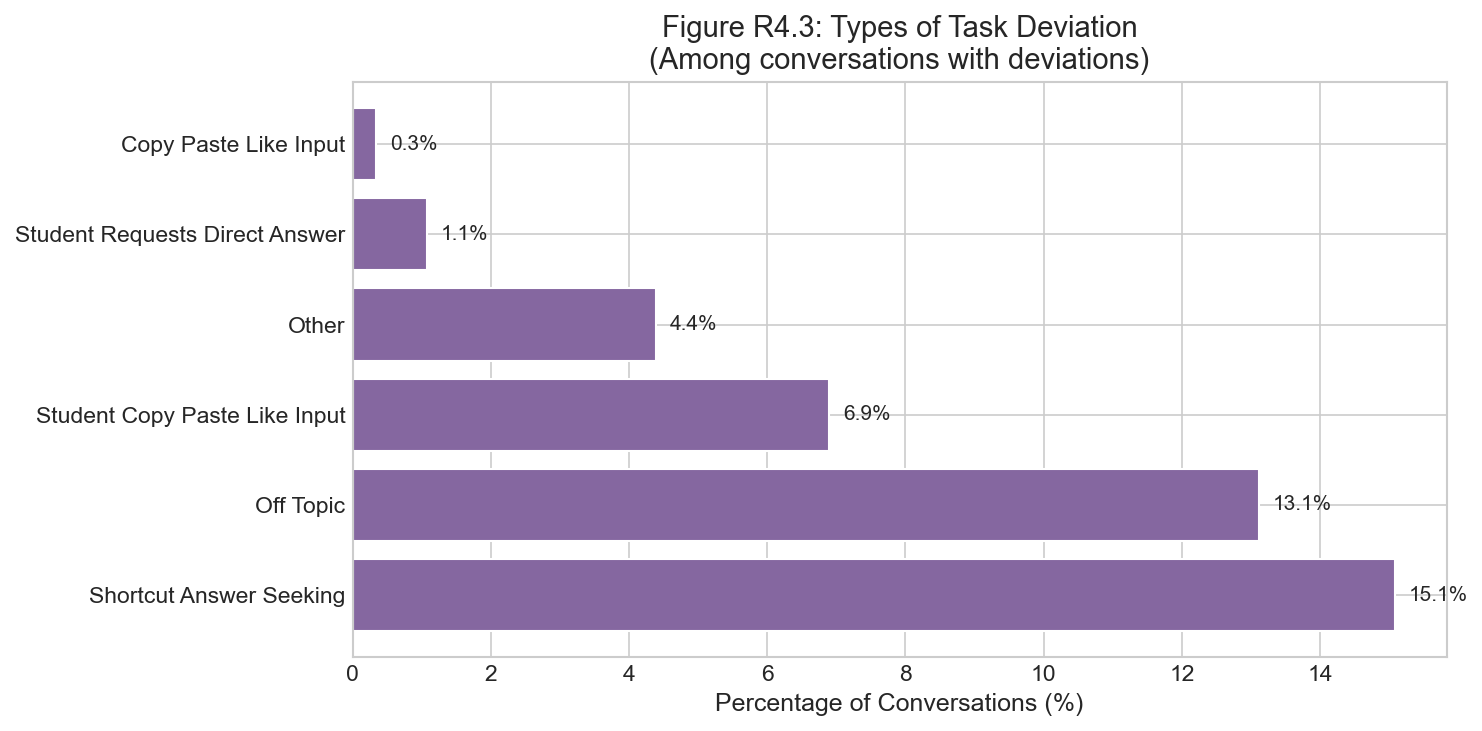}
\caption{Types of task deviation among conversations with alignment issues. Shortcut/answer-seeking and off-topic exchanges account for the majority of partial deviations, while substantial off-task behavior is rare.}
\label{fig:deviation_types}
\end{figure}

\begin{figure}[ht]
\centering
\includegraphics[width=\columnwidth]{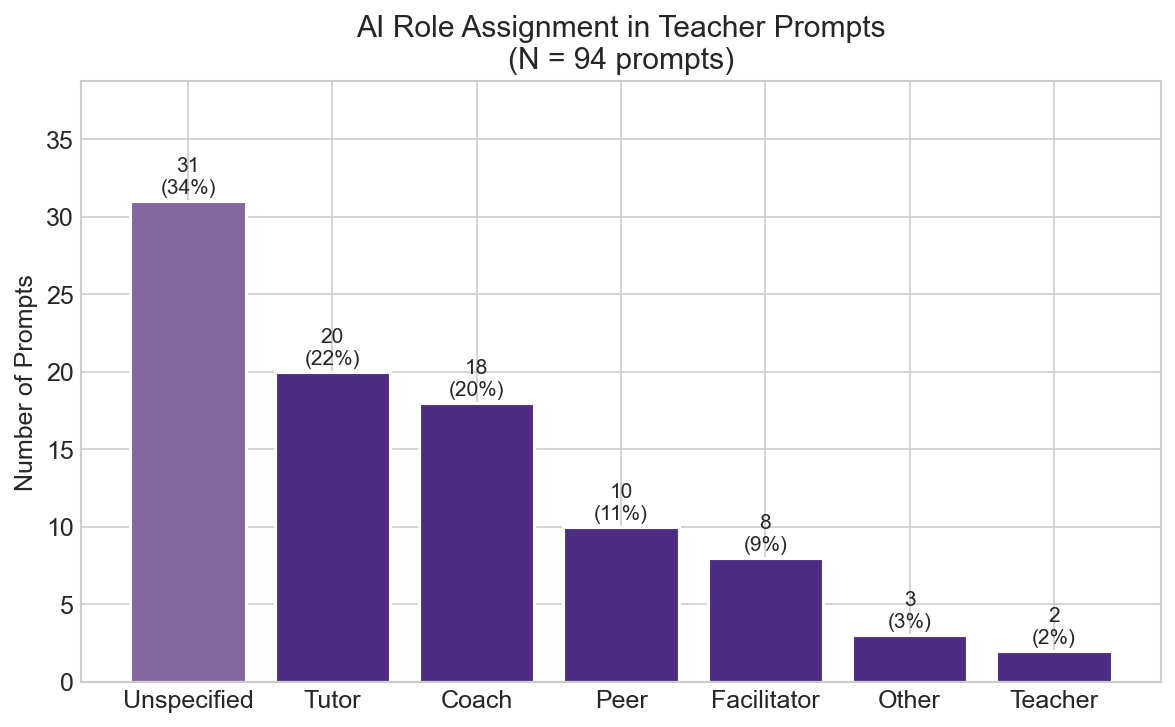}
\caption{Distribution of AI role assignments in teacher prompts.}
\label{fig:roles}
\end{figure}

\begin{figure}[ht]
\centering
\includegraphics[width=\columnwidth]{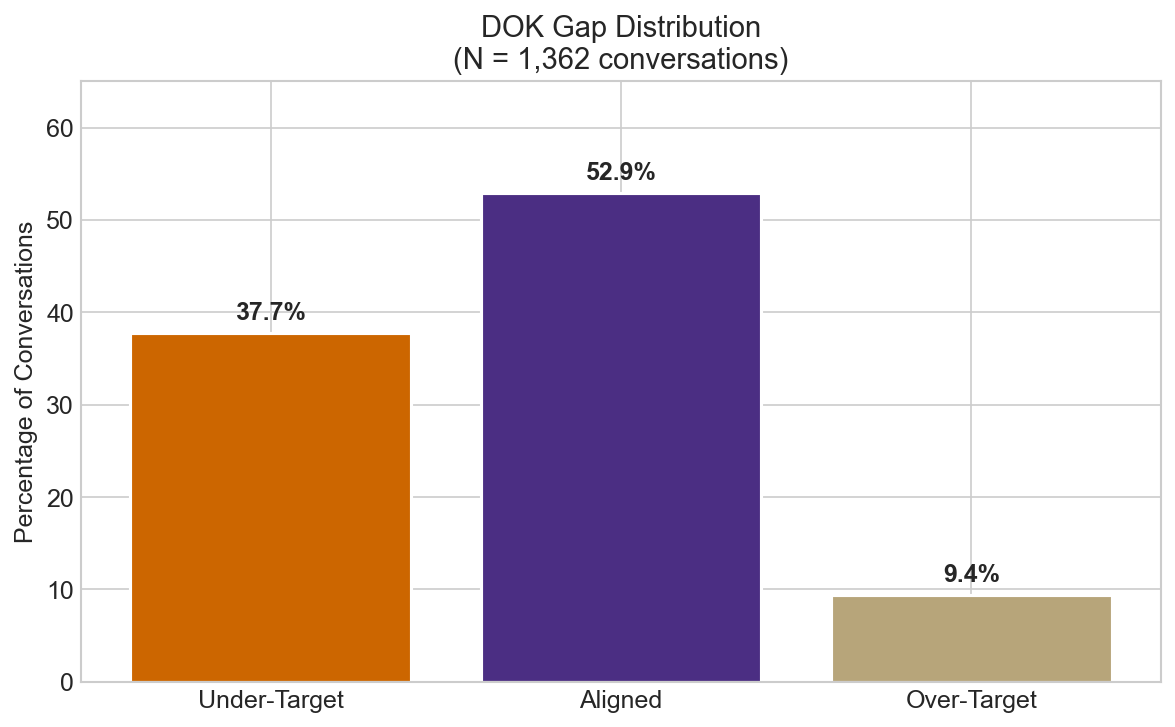}
\caption{Distribution of DOK gap categories across conversations (N = 1,362). Positive values indicate under-targeting.}
\label{fig:dok_gap}
\end{figure}

\begin{figure}[ht]
\centering
\includegraphics[width=\columnwidth]{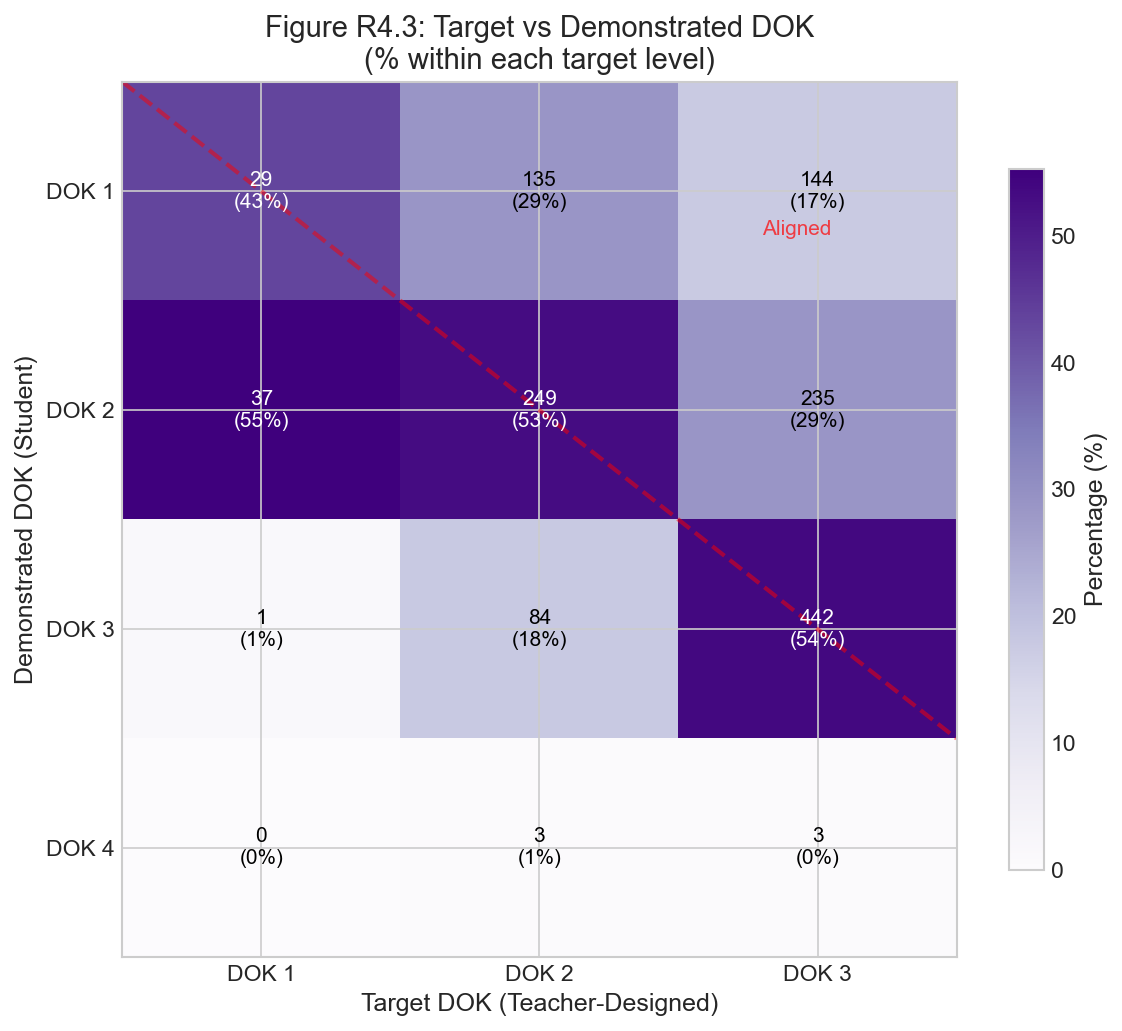}
\caption{Heatmap of target vs. demonstrated DOK levels. The diagonal represents perfect alignment; cells above the diagonal indicate under-targeting.}
\label{fig:dok_heatmap}
\end{figure}

\begin{table}[ht]
\centering
\caption{AI behavioral patterns across conversations}
\label{tab:ai_behaviors}
\begin{tabular}{lr}
\toprule
\textbf{AI Behavior} & \textbf{\% of Conversations} \\
\midrule
Asks Socratic questions & 90.4\% \\
Provides stepwise guidance & 72.1\% \\
Requests student attempt first & 70.5\% \\
Requests evidence or reasoning & 69.0\% \\
Refuses/limits direct answers & 45.8\% \\
Provides hints first & 40.8\% \\
Provides final answer/solution & 14.7\% \\
\textbf{Explicit safeguard violation} & \textbf{0.0\%} \\
\bottomrule
\end{tabular}
\end{table}

\begin{figure}[ht]
\centering
\includegraphics[width=\columnwidth]{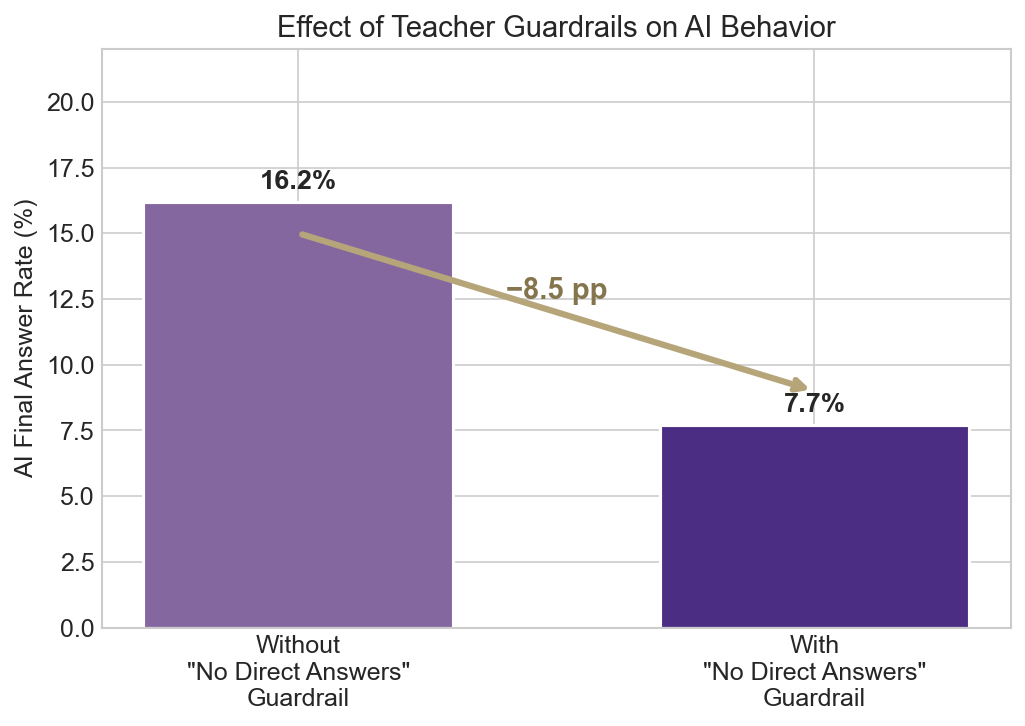}
\caption{Effect of teacher ``no direct answers'' guardrail on AI final answer rate.}
\label{fig:guardrail_effect}
\end{figure}

\section{Extended Qualitative Findings}
\label{app:qual_results}

Qualitative data from semi-structured interviews ($N=10$) and teacher reflections contextualized the trace-based patterns reported above. Across sources, teachers framed TASD as the introduction of a `third agent'' that could support individualized instruction and feedabck, but only when the activity was developmentally appropriate, tightly framed, and supported by teacher-facing visibility in fast-paced real-time fashion. First, teachers emphasized \textit{developmental fit and cognitive load} as a determinant of engagement. When the AI relied on probing questions without an immediately actionable next step, or when responses were overly verbose, some students disengaged. This pattern got further exacerbated when AI responses were long with multiple follow-up questions. Teachers also described this as an enactment problem. When the interaction demanded substantial interpretation, it competed with increased the students' burden to translate AI moves into next steps they could execute. Second, teachers described \textit{within-class heterogeneity} in participation, often observing `all-or-nothing'' patterns in which some students sustained long exploratory exchanges while others contributed minimally or opted out. Teachers attributed this variability to differences in student readiness, perceptions of AI's effectiveness and broader social implications, classroom norms, and the clarity of the deliverable. This aligns with the quantitative finding that better-designed prompts, such as explicit finish lines, are associated with reduced rigor gaps, while also pointing to an unobserved layer of implementation--what unfolds during classroom enactment, including students' prior AI exposure and literacy, teachers framing, and students' learning styles. Teachers reported that some students refused to use AI due to concerns about its environmental impact or fears that teachers would offload work to AI and reduce direct interaction with students. At the same time, teachers described meaningful benefits for some learners, especially students who felt less confident in peer discussion and advanced students who sought deeper intellectual stimulation. For example, one eighth-grade mathematics teacher noted that multilingual learners who previously resisted speaking in front of the class felt more comfortable contributing after practicing and rehearsing with the AI. Third, teachers identified \textit{visibility and monitoring} as prerequisites for viability. Teachers described early implementations as difficult to supervise without lightweight indicators of who had started, who was stuck, and who had completed the task. Teachers reported that monitoring supports reduced orchestration burden by enabling rapid triage and targeted intervention during live sessions. Nevertheless, teachers still needed to review student-AI transcripts across the class to ensure appropriate engagement and integrity. For example, teachers described cases in which students copied and pasted answers obtained elsewhere (e.g., from ChatGPT) after failing to elicit direct answers from TASD.

\section{A taxonomy of teacher-authored TASD use cases}
\label{app:use cases}
Teachers used TASD not as a generic chatbot but as a configurable instructional routine embedded in classroom activity. We summarize the dominant use-case clusters below, illustrated by pilot artifacts (paraphrased to avoid reproducing full teacher text). Across up to $\sim$600 participating students, TASD supported repeated in-class activities spanning discussion preparation, SEL routines, language scaffolding, and role-play. Although these clusters are presented as distinct, teachers frequently combined them (e.g., MLL constraints plus discussion preparation; role-play plus rubric-based self-check), indicating that authoring centered on assembling a small set of reusable interaction patterns into lesson-specific routines.

\paragraph{U1. Discussion preparation and text-based sensemaking (ELA/SS).}
Teachers configured TASD to help students prepare for whole-class discussion by prompting theme identification, vocabulary support, and evidence-based reasoning from a shared text. These designs often included explicit norms (e.g., ``use evidence''), structured prompts, and a clear finish line (e.g., generate 2--3 evidence-backed ideas before discussion).

\paragraph{U2. Multilingual learner (MLL) micro-scaffolds and accessibility constraints.}
A prominent pattern was short-turn interaction (e.g., ``respond with one sentence'' / ``ask one question'') paired with simplified language for students with emerging proficiency. Teachers used TASD for low-stakes check-ins, comprehension repair, and guided reflection with constrained cognitive load. This aligns with broader evidence that chatbot effectiveness depends on interaction design and learner characteristics.

\paragraph{U3. SEL and classroom routines (check-ins, gratitude, goal setting).}
Teachers authored short, low-stakes dialogues to build emotional awareness, gratitude, and goal-setting routines (e.g., Monday check-in, Friday reflection). These were often used as entry tasks, emphasizing brevity and psychological safety.

\paragraph{U4. Formative practice toward mastery with explicit criteria.}
Some teachers configured iterative practice loops tied to a target skill level (e.g., cause--effect reasoning) with repeated attempts until students met a performance descriptor. While not a formal assessment, these designs used mastery framing and frequent checks, consistent with formative feedback principles.

\paragraph{U5. Metacognitive coaching and learning strategy dialogue.}
Teachers used TASD to prompt students to explain learning strategies (e.g., why note-taking supports memory), often requiring the student to steer the dialogue while the AI stayed concise. These designs aim to externalize reasoning and support self-regulated learning behaviors.

\paragraph{U6. Project management and accountability supports.}
Teachers configured dialogues that chunked tasks (e.g., ``one question at a time'') to help students plan progress, self-assess against a rubric, and reflect on collaboration contributions. These designs treated TASD as a coach that reduces overwhelm while maintaining student ownership.

\paragraph{U7. Schema/background knowledge building for inaccessible texts.}
When texts were not accessible at students’ reading levels, teachers used TASD to build background knowledge and vocabulary before encountering the text, supporting comprehension and participation in subsequent reading or discussion.

\paragraph{U8. Literacy subskills: morphology and vocabulary categorization.}
Teachers built structured routines for phoneme/morpheme breakdown and for grouping vocabulary by category, with error correction and re-tries. These were tightly constrained interactions with clear correctness criteria.

\paragraph{U9. Role-play simulations and club/CTE preparation.}
Several TASDs were configured as role-based tutors (e.g., mock interviews, DECA event preparation, business/finance consultant scenarios) using modes such as explore, scenario, and quiz. These designs leveraged conversational affordances for practice in applied contexts.

\end{document}